\documentclass[runningheads,natbib]{svjour2}

\usepackage[figuresright]{rotating}
\usepackage{longtable}
\usepackage{graphicx}
\usepackage{url}
\newcommand{\kms}{{\,\rm km\,s}^{-1}} 
 
\newcommand{\nodata}{\multicolumn{1}{c}{$\cdots$}}

\def\la{\mathrel{\hbox{\rlap{\hbox{\lower4pt\hbox{$\sim$}}}\hbox{$<$}}}}

\renewcommand{\mag}{\mbox{$\;$mag}}
\def\aap{A\&A }
\def\aaps{A\&A Suppl. }
\def\aj{Astron. J. }
\def\apj{ApJ }
\def\apjs{ApJ Suppl. }

\def\mnras{MNRAS }

\def\pasp{Publ. Astron. Soc. Pacific }

\def\araa{ARA\&A }

\journalname{Astronomy and Astrophysics Review}
\begin{document}
% ******************************************************************
\title{The expansion field: The value of \boldmath{$H_{0}$}}
\titlerunning{The expansion field: The value of $H_{0}$}   

\author{G. A. Tammann         \and
        A. Sandage            \and
        B. Reindl
}
\authorrunning{Tammann, Sandage \& Reindl}

\institute{G.\,A. Tammann, B. Reindl \at 
           Department of Physics and Astronomy,\\
           Klingelbergstrasse 82, CH-4056 Basel, Switzerland.
           \email{g-a.tammann@unibas.ch}
           \and
           A. Sandage \at
           Observatories of the Carnegie Institution of Washington,\\
           813 Santa Barbara Street, Pasadena, CA 91101
}

% \date{\today}
\date{\vskip-7em\noindent\today} % ... to move keywords on
                                 % titlepage for preprint

\maketitle

% ******************************************************************
%    Abstract
% ******************************************************************
\begin{abstract}
Any calibration of the present value of the Hubble constant ($H_{0}$)
requires recession velocities and distances of galaxies. While the
conversion of observed velocities into true recession velocities has
only a small effect on the result, the derivation of unbiased
distances which rest on a solid zero point and cover a useful range of
about 4 to 30$\;$Mpc is crucial.
A list of 279 such galaxy distances within $v<2000\kms$ is given which
are derived from the tip of the red-giant branch (TRGB), from
Cepheids, and/or from supernovae of type Ia (SNe\,Ia). 
Their random errors are not more than $0.15\mag$ as shown
by intercomparison. They trace a {\em linear\/} expansion field within
narrow margins, supported also by external evidence, from $v=250$ 
to at least $2000\kms$. Additional 62 distant SNe\,Ia confirm the
linearity to at least $20,000\kms$. The dispersion about the Hubble
line is dominated by random peculiar velocities, amounting locally to
$<100\kms$ but increasing outwards. Due to the linearity of the
expansion field the Hubble constant $H_{0}$ can be found at any
distance $>4.5\;$Mpc. 
RR\,Lyr star-calibrated TRGB distances of 78 galaxies above this limit
give $H_{0}=63.0\pm1.6$ at an effective distance of $6\;$Mpc. 
They compensate the effect of peculiar motions by their large number. 
Support for this result comes from 28 {\em independently\/} calibrated
Cepheids that give $H_{0}=63.4\pm1.7$ at 15 Mpc. 
This agrees also with the large-scale value of $H_{0}=61.2\pm0.5$ from
the distant, Cepheid-calibrated SNe\,Ia. A mean value of 
$H_{0}=62.3\pm1.3$ is adopted. Because the value depends on two
independent zero points of the distance scale its systematic error is
estimated to be 6\%. -- 
Other determinations of $H_{0}$ are discussed. 
They either conform with the quoted value (e.g.\ line width data of
spirals or the $D_{n}\!-\!\sigma$ method of E galaxies) or are judged
to be inconclusive. Typical errors of $H_{0}$ come from the use of a
universal, yet unjustified P-L relation of Cepheids, the neglect of
selection bias in magnitude-limited samples, or they are inherent to
the adopted models.   
\end{abstract}
% ******************************************************************

% ******************************************************************
% 1. Introduction
% ******************************************************************
\section{Introduction}
\label{sec:1}
It is said sometimes that once in a career, every astronomer is
entitled to write a paper on the value of the Hubble constant. 
To the point, several compilations of the growing literature on
$H_{0}$ since 1970 have been made. Those by \citet{Press:97}, 
\citet{Tammann:06} and \citet{Huchra:07} are examples.      

     These authors plot histograms of the distribution of $H_{0}$ from
about 400 papers since 1970. The sample is so large that the formal
error on the average of the histogram is so small that one might infer
that the Hubble constant is now known to better than say 1\%.     
Of course, what is missing is the fact that most of the values
in the literature are not correct. Many suffer from the neglect of the
effects of an observational selection bias that {\em varies with
distance}. 

     We are faced with a problem in writing this review. Do we strive
to give a comprehensive history of the distance scale problem
beginning with the first determination of the Hubble constant by 
\citet{Lemaitre:27,Lemaitre:31,Robertson:28,Hubble:29b}, 
and \citet{Hubble:Humason:31,Hubble:Humason:34} to be about 
$550\;\mbox{km}\;\mbox{s}^{-1}\;\mbox{Mpc}^{-1}$
(units assumed hereafter), coming into modern times with the debates
between the principal players? Or do we only write about the situation
as it exists today, comparing the ``concordance'' value of $H_{0}=72$
by \citet{Freedman:etal:01} with the {\em HST\/} supernovae
calibration value (\citealt{Hamuy:etal:96}; \citealt{Tripp:Branch:99};
\citealt{Suntzeff:etal:99}; 
\citealt{STT:06}, hereafter \citeauthor*{STT:06}; 
\citealt{STS:06}, hereafter \citeauthor*{STS:06})
that gives $H_{0}=62$? We have
decided to take the latter course but also to sketch as a skeleton the
beginning of the correction to Hubble's 1930-1950 distance scale that
started with the commissioning of the 200-inch telescope in 1949. An
important comprehensive review of this early period before the Hubble
Space Telescope ({\em HST}) is by \citet{Rowan-Robinson:85}; the
details are not repeated here.

% ********************************************************
% 1.1. Early work on the revision to Hubble's distance scale (1950-1990)
% ********************************************************
\subsection{Early work on the revision to Hubble's distance scale (1950-1990)}
\label{sec:01:1}
Hubble's extragalactic distance scale was generally believed from
1927 to about 1950, beginning with the first determinations of the Hubble
constant by the four independent authors cited above.
This scale lasted until Hubble's \citeyearpar{Hubble:29a} distance to
M\,31 was nearly tripled by \citet{Baade:54} in his report to the 1952
Rome meeting of the IAU. He proposed a revision of the Cepheid P-L
relation zero point by about $1.5\mag$ based on his discovery that 
RR\,Lyrae stars could not be detected with the newly commissioned
200-inch Palomar telescope in M\,31 at $V=22.2$. From this he
concluded that M\,31 must be well beyond the modulus of $(m-M)=22.2$
given earlier by Hubble. The story is well known and is recounted
again by \citet[][Chapter~6]{Osterbrock:01}, in the Introduction to
\citet{TSR:08}, hereafter \citeauthor*{TSR:08}, and often in histories
elsewhere \citep[e.g.][]{Trimble:96,Sandage:99a} .    

     Following Baade's discovery, the revision of 1930-1950 scale was
begun anew with the Palomar 200-inch telescope, largely following
\citet*{Hubble:51} proposed cosmological program for it.  
Observational work on the first Cepheid distance beyond the Local 
Group was completed for NGC\,2403. Here we made photoelectric
calibrations of magnitude scales and used new calibrations of the
Cepheid P-L relations \citep{Kraft:61,Kraft:63,ST:68,ST:69},
and we obtained a revised distance modulus of $(m-M)=27.56$
\citep{TS:68}. Comparing this with Hubble's modulus of 24.0 showed the
large scale difference by a factor of 5.2. 
Next, the modulus of the more remote galaxy, M\,101, was 
determined to be $(m-M) = 29.3$ \citep{ST:74a} 
compared with Hubble's modulus of 24.0, giving the large 
correction factor of 11.5 to Hubble's scale at M\,101 
($D = 7.2\;$Mpc). This large stretching was again found in our
distance modulus of $(m-M)=31.7$ for the Virgo cluster 
\citep{ST:74b,ST:76,ST:90,ST:95}, compared with Hubble's modulus of
26.8. The distance ratio here is a factor of 9.6.        

     These large factors and their progression with distance came as a
major shock in the mid 1970s and were not generally believed
\citep[eg.][etc.]{Madore:76,Hanes:82,deVaucouleurs:82}.  
However, the new large distances were confirmed for NGC\,2403 by 
\citet{Freedman:Madore:88}, and for M\,101 by \citet{Kelson:etal:96}
and \citet{Kennicutt:etal:98}. Although our distance to the Virgo 
cluster core is still in contention at the 20\% level, there is no
question that the correction factor here is also between 7 and 10 at
$20\;$Mpc.

% ********************************************************
% 1.2. The difficulty of finding H_0
% ********************************************************
\subsection{The difficulty of finding $H_{0}$}
\label{sec:01:2}
The determination of $H_{0}$, the {\em present\/} and hence nearby
value of the Hubble parameter, requires -- besides true recession
velocities -- distance indicators with known zero point {\em and\/}
with known intrinsic dispersion. The scatter of the Hubble diagram,
$\log v$ versus $m$ or $(m-M)$, would in principle be a good
diagnostic for the goodness of a given distance indicator if it were
not also caused by peculiar motions. It is of prime importance to
disentangle these two sources of scatter because unacknowledged
intrinsic scatter of the available distances introduces a 
{\em systematic\/} increase of $H_{0}$ with distance if {\em
  flux-limited samples\/} are considered, which is normally the
case. This is because the mean absolute magnitude of objects in such
samples increases with distance due to the increasing discrimination
against the less luminous objects. 
It is important to note that, strictly speaking, this incompleteness
bias is {\em not\/} the \citet{Malmquist:20,Malmquist:22} bias which
applies only to the {\em average\/} effect integrated over the sample
being studied; not to individual distances within that sample, each of
which must be corrected by a sliding scale. 

     Neglect of the individual bias values that become
progressively larger with increasing distance always gives a Hubble
constant that incorrectly appears to increase outward 
\citep{deVaucouleurs:58,deVaucouleurs:76,deVaucouleurs:77,Tully:88}. 

     The widely held view that the increase of $H_{0}$ with
distance (up to an unspecified limit) was real deprived the Hubble
diagram of its second diagnostic power. The slope of the Hubble line
had no longer to be $0.2$, which is the case for {\em linear\/}
expansion (see hereafter Eq.~\ref{eq:01}). The apparent increase of
$H_{0}$ with distance was not anymore accepted as proof for bias  
(e.g.\ \citealt{Tammann:87} versus \citealt{Aaronson:87}).
It also led to proposals that $H_{0}$ not only varied with distance,
but also with direction 
\citep{deVaucouleurs:Bollinger:79,deVaucouleurs:Peters:85}.
The search for the asymptotic value of $H_{0}$ became self-defeating:
one tried to calibrate it at the largest possible distances where,
however, the effects of bias are largest.

     The bias is always present in a flux limited sample of field
galaxies \citep[][as analyzed using Spaenhauer diagrams]{%
Sandage:94a,Sandage:94b,Sandage:95,FST:94}.
It is also present in cluster data that are incomplete 
\citep{Teerikorpi:87,Teerikorpi:90,Kraan-Korteweg:etal:88,
Fouque:etal:90,STF:95,Sandage:08},
and even in field galaxies of any sample that {\em is\/} distance
limited but if the data are incomplete in the coverage of the distance
indicator (apparent magnitude, 21cm line width, etc.)
\citep{Sandage:08}. 

     However, claims for $H_{0}$ increasing outwards were contradicted
by the apparent magnitudes of first-ranked galaxies in clusters and
groups. The Hubble diagram of brightest cluster galaxies shows no
deviations from linear expansion down to $\sim\!2000\kms$ 
\citep[][and references therein]{STH:72,Sandage:Hardy:73,Kristian:etal:78}.
This was confirmed down to $\sim\!1000\kms$  in a study of northern
and southern groups \citep{Sandage:75}, which also showed a smooth
linear Hubble diagram with no discontinuities over the range of
$1000<v<10,000\kms$. The limit on $\delta H_{0}/H_{0}$ was $<0.08$,
and a proof was given that the Hubble constant does not increase
outward. These results were confirmed by \citet{FST:94}
based on the large catalog of 21cm line widths and $I$ magnitudes by
\citet{Mathewson:etal:92a,Mathewson:etal:92b}, and also in the large
archive literature cited therein by many others. 
However, it was so far not possible to tie the local expansion field
below $\la15\;$Mpc into the large-scale field because of small-number
statistics and of large scatter caused by the important effects of
peculiar velocities and distance errors. This problem is the subject
of Sect.~\ref{sec:2}.

     In parallel to the discussion on distance errors there were many
attempts to determine the mean size of the random
one-dimensional peculiar velocities $v_{\rm pec}$ by reading the
deviations from the Hubble line vertically as velocity residuals, but
this is not easier than to determine the dispersion of the distance
indicators because the latter have to be known. In fact the problem is
here even deeper. The halted expansion of the Local Group, the
retarded expansion by the gravity of the Virgo complex, the large
virial velocities in clusters, and the increase of peculiar motions
with distance, as manifested by the important velocity of a large
volume with respect to the CMB dipole all make it difficult to find
the characteristic peculiar velocities of field galaxies.

     One of the earliest attempts to determine a cosmological
parameter of interest (other than $H_{0}$) was that by Hubble and
Humason to measure the mean random velocity of galaxies about an
ideal Hubble flow. This, in turn, is related to any systematic
streaming, or more complicated systematic motions (a dipole plus even 
a quadrupole, a shear, or a local rotation) relative to a cosmic frame  
(\citealt{Davis:Peebles:83a} for a review; see also
\citealt{Dekel:94}).  
The discussion by \citet{Hubble:Humason:31} gave values between 
$200$ and $300\kms$ for the mean random motion (they do not 
quote an rms value) about the ridge line of the redshift-distance
relation for local galaxies ($v < 10,000\kms$). 

     By 1972 a limit was set of $v_{\rm pec}<100\kms$ on local scales
\citep{Sandage:72}. In subsequent papers, too numerous to be cited
here, rather lower values were favored
\citep[e.g.][]{ST:75,Giraud:86,Sandage:86,Ekholm:etal:01,Thim:etal:03}.
In a representative study \citet{Karachentsev:Makarov:96} found 
$v_{\rm pec}=72\kms$, supported by later papers of Karachentsev and
collaborators. The values of $v_{\rm pec}$ in function of scale length
agree locally (see Sect.~\ref{sec:2:5}), but clearly increase with
distance. 

     The modest size of the peculiar velocities poses a problem for
various hierarchical merging scenarios of galaxy formation which
predict mean random motions as high as $500\kms$   
\citep[cf.][]{Davis:Peebles:83b,Davis:etal:85,Ostriker:93,Governato:etal:97,Leong:Saslaw:04}.

% ******************************************************************
% 2. The local expansion field
% ******************************************************************
\section{The local expansion field}
\label{sec:2}
The search for the cosmic (global) value of the Hubble constant $H_{0}$
requires some a priori knowledge of the expansion field. How linear is
the expansion? Does $H_{0}$ vary with distance? How large are typical
peculiar motions and/or streaming velocities which may lead to
incorrect results on $H_{0}$? Only once these questions are answered
it is possible to judge the goodness of other distance indicators by
the shape and the tightness of their Hubble diagrams. While a detailed
mapping of non-Hubble motions in function of individual density
fluctuations is important in its own right, it is not necessary here.
For the average value of $H_{0}$ from an all-sky sample of galaxies it
is enough to know the dependence of $H_{0}$ on distance over scales of
$\ge3\;$Mpc as well as the effect of peculiar motions on the available
sample. The problem of large virial motions in clusters can be
circumvented by assigning the mean cluster velocity to individual
members.

     Mapping the expansion field requires hence a significant number
of {\em relative\/} distances with a sufficient range and with minimum
intrinsic scatter to guard against selection effects which distort
the field. Even in case of more than one distance indicator used for
the mapping, only relative distances are needed because they can be 
combined by requiring that they obey the same expansion rate $H_{0}$
within a given distance range, i.e.\ that they have the same intercept
$a$ of the Hubble diagram. Note that
\begin{equation}
   \log v = 0.2 m^{0}_{\lambda} + C_{\lambda}, \quad \mbox{where}
\label{eq:01}
\end{equation}
\begin{equation}
   C_{\lambda} =  \log H_{0} -  0.2 M^{0}_{\lambda} - 5.
\label{eq:02}
\end{equation}
($m^{0}_{\lambda}$ is the apparent, absorption-corrected magnitude of
a galaxy at wavelength $\lambda$; $M^{0}_{\lambda}$ is the
corresponding absolute magnitude).
In case that the mean absolute magnitude is assumed to be known or
that the true distance moduli are known this becomes 
\begin{equation}
   \log v = 0.2 (m-M)^{0} + a, \quad \mbox{from which follows}
\label{eq:03}
\end{equation}
\begin{equation}
   \log H_{0} = a + 5. 
\label{eq:04}
\end{equation}

     Many data have become available during the last years for three
distance indicators that are ideally suited for the purpose of
expansion field mapping because they provide distance moduli with
random errors of only $\le 0.15\mag$ (corresponding to 7.5\% in
distance) as shown in Section~\ref{sec:3} by intercomparison. 
These distance indicators are the tip of the red-giant branch (TRGB),
classical Cepheids, and supernovae of type Ia at maximum
luminosity (SNe\,Ia). Table~\ref{tab:dist} below lists 240 TRGB, 43
Cepheid, and 22 SNe\,Ia distances outside the Local Group, which
provide the backbone of the determination of $H_{0}$. 

     Although relative distances are all that is needed to test the
{\em linearity\/} of the expansion field and its peculiar motions,
absolute distances as zero-pointed in Section~\ref{sec:3} will be used
in the following simply because they are available. This has the
advantage that differences of the intercept $a$ of the particular
Hubble diagrams yield an estimate of the systematic error of the
adopted distance scale.

% ********************************************************
% 2.1 Corrections of the distances and of the velocities
% ********************************************************
\subsection{Corrections of the distances and of the velocities}
\label{sec:2:1}
{\em All distances in this paper} (outside the Local Group) 
{\em are transformed to the barycenter of the Local Group\/} 
which is assumed to lie at the distance of $0.53\;$Mpc in the
direction of M\,31, i.e.\ at two thirds of the way to this galaxy,
because the galaxies outside the Local Group expand presumably away
from the barycenter and not away from the observer. 
Distance moduli from the observer, corrected for Galactic absorption,
are designated with $\mu^{0}\equiv(m-M)^{0}$, while $\mu^{00}$ stands
for the moduli reduced to the barycenter.

     The heliocentric velocities $v_{\rm hel}$ are corrected to the
barycenter of the Local Group following \citet{Yahil:etal:77} and -- 
except for Local Group galaxies -- for a self-consistent Virgocentric
infall model assuming a local infall vector of $220\kms$ and a density
profile of the Virgo complex of $r^{-2}$
(\citealt{Yahil:etal:80},
\citealt{Dressler:84},
\citealt{Kraan-Korteweg:86},
\citealt{deFreitasPacheco:86},
\citealt{Giraud:90},
\citealt{Jerjen:Tammann:93},
see Eq.~(5) in \citeauthor*{STS:06}). 
The choice of these particular corrections among others proposed in
the literature is justified because they give the smallest scatter in
the Hubble diagrams (\citeauthor*{STS:06}). 
Velocities relative to the barycenter are designated with $v_{0}$;
velocities corrected for Virgocentric infall (which makes of course no
sense for members of the bound Local Group) are designated with
$v_{220}$. -- 
The velocities of galaxies outside the Local Group are also corrected
for the projection angle between the observer and the Local Group
barycenter as seen from the galaxy, but the correction is negligible
except for the very nearest galaxies. 

     The Virgocentric infall corrections are only a first
approximation. The actual velocity field is much more complex as seen
in the model of \citet{Klypin:etal:03}. But any such corrections have
surprisingly little influence on the all-sky value of $H_{0}$ even at
small distances (Sect.~\ref{sec:3:4:2}). The main effect of the
adopted infall-corrected $v_{220}$ velocities is that they yield a
noticeably smaller dispersion of the Hubble diagram, as stated before,
than velocities which are simply reduced to the barycenter of the Local 
Group.

     Galaxies with $v_{0}>3000\kms$ are in addition corrected for the
CMB dipole motion on the assumption that the comoving local volume
extends out to this distance \citep{FST:94}. 
Even if the merging into the background field kinematics takes place
as far out as $6000\kms$ \citep{Dale:Giovanelli:00} it has no
noticeable effect on the present conclusions.

% ********************************************************
% 2.2 The Hubble diagram of TRGB distances
% ********************************************************
\subsection{The Hubble diagram of TRGB distances}
\label{sec:2:2}
\begin{figure}
   \centering
   \includegraphics[width=0.79\textwidth]{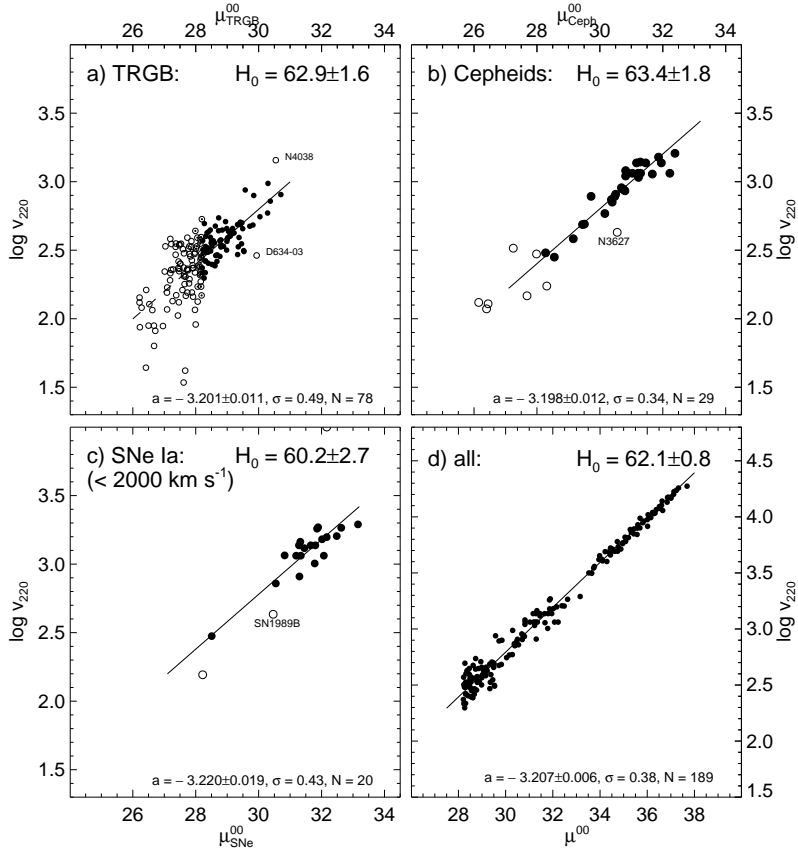}
    \caption{The Hubble diagram of 
    a) TRGB, b) Cepheids, and c) SNe\,Ia cut at
    $v_{220}<2000\kms$. Panel d) shows all galaxies of a) $-$ c) plus
    the SNe\,Ia with $v_{\rm CMB}<20,000\kms$.}
    \label{fig:hubble}
\end{figure}

The galaxies outside the Local Group with available TRGB distances are 
listed in Table~\ref{tab:dist}.
The identifications of the galaxies in Col.~1 are from the NED (NASA
Extragalactic Database, http://nedwww.ipac.caltech.edu/index.html); in
some cases they are here slightly abbreviated. Alternative designations
are given in the same source. The group assignments in Col.~2 are
evaluated from various sources. The heliocentric velocities in Col.~3
are from the NED. The distances $\langle \mu^{0}\rangle$ in Col.~9 are
the straight mean of the available distance determinations as seen
from the observer. Col.~10 gives the mean distances 
$\langle \mu^{00}\rangle$ reduced to the barycenter of the Local
Group. The latter are plotted in a Hubble diagram 
(Fig.~\ref{fig:hubble}a).  
The 78 galaxies with distances $>4.4\;$Mpc and up to $\sim10\;$Mpc
yield a free-fit Hubble line with slope $0.166\pm0.019$ if $\log
v_{220}$ is used as the independent variable, and with slope 
$0.332\pm0.038$ if $\mu^{00}$ is used as the independent variable. 
The orthogonal solution, i.e.\ the mean of the two previous solutions,
gives a slope of $0.199\pm0.019$, which is so close to 0.2 that a
forced fit with slope 0.2 is justified even for this very local
volume.  

     The dispersion in Fig.~\ref{fig:hubble}a, read in $\mu^{00}$, is
$\sigma_{\mu}=0.49$. This value rests mainly on the effect of peculiar
motions. The random error of the distances is not more than $0.15\mag$
(Sect.~\ref{sec:3:4:1}). Also observational errors of the velocities
contribute little to the dispersion. Hence the contribution of the
peculiar motions must be close to $0.47\mag$.  

     A still closer sample of 20 TRGB galaxies in Table~\ref{tab:dist}
within the narrow distance interval $3.9-4.4\;$Mpc can of course not
provide a test for the slope. Yet assuming a slope of 0.2 gives the
same intercept $a$ and hence the same mean Hubble constant as from the
more distant TRGB distances to within 5\%.  
The dispersion of this nearby sample is large at
$\sigma_{\mu}=0.74$. It may be increased by observational
velocity errors, which for some dwarf galaxies may amount
to $\sim\!50\kms$. Therefore the contribution of the peculiar
velocities is here not well determined.

% ***********************************************
%  Table 1: High accuracy distances of local galaxies
% ***********************************************
\scriptsize
\begin{longtable}[l]{llrrccccccc}
\caption{High accuracy distances of local galaxies.\label{tab:dist}}\\[-6pt]
% ***********************************************
\hline
\hline
\noalign{\smallskip}
% ***********************************************
\multicolumn{1}{c}{Galaxy}                  & 
\multicolumn{1}{c}{Group}                   & 
\multicolumn{1}{c}{$v_{\rm hel}$}           &
\multicolumn{1}{c}{$v_{220}$}               &
\multicolumn{1}{c}{$\mu^{0}_{\rm RRLyr}$}   &
\multicolumn{1}{c}{$\mu^{0}_{\rm TRGB}$}    &
\multicolumn{1}{c}{$\mu^{0}_{\rm Cep}$}     &
\multicolumn{1}{c}{$\mu^{0}_{\rm SNe}$}     &
\multicolumn{1}{c}{$\langle\mu^{0}\rangle$} &
\multicolumn{1}{c}{$\langle\mu^{00}\rangle$}&
\multicolumn{1}{c}{Ref} \\
\multicolumn{1}{c}{(1)}                     & 
\multicolumn{1}{c}{(2)}                     & 
\multicolumn{1}{c}{(3)}                     & 
\multicolumn{1}{c}{(4)}                     & 
\multicolumn{1}{c}{(5)}                     & 
\multicolumn{1}{c}{(6)}                     & 
\multicolumn{1}{c}{(7)}                     & 
\multicolumn{1}{c}{(8)}                     & 
\multicolumn{1}{c}{(9)}                     & 
\multicolumn{1}{c}{(10)}                    & 
\multicolumn{1}{c}{(11)}                    \\ 
% ***********************************************
\noalign{\smallskip}
\hline
\noalign{\smallskip}
\endfirsthead
% ***********************************************
\caption{(Continued)}\\[-6pt]
\hline
\hline
\noalign{\smallskip}
% ***********************************************
\multicolumn{1}{c}{Galaxy}                  & 
\multicolumn{1}{c}{Group}                   & 
\multicolumn{1}{c}{$v_{\rm hel}$}           &
\multicolumn{1}{c}{$v_{220}$}               &
\multicolumn{1}{c}{$\mu^{0}_{\rm RRLyr}$}   &
\multicolumn{1}{c}{$\mu^{0}_{\rm TRGB}$}    &
\multicolumn{1}{c}{$\mu^{0}_{\rm Cep}$}     &
\multicolumn{1}{c}{$\mu^{0}_{\rm SNe}$}     &
\multicolumn{1}{c}{$\langle\mu^{0}\rangle$} &
\multicolumn{1}{c}{$\langle\mu^{00}\rangle$}&
\multicolumn{1}{c}{Ref} \\
\multicolumn{1}{c}{(1)}                     & 
\multicolumn{1}{c}{(2)}                     & 
\multicolumn{1}{c}{(3)}                     & 
\multicolumn{1}{c}{(4)}                     & 
\multicolumn{1}{c}{(5)}                     & 
\multicolumn{1}{c}{(6)}                     & 
\multicolumn{1}{c}{(7)}                     & 
\multicolumn{1}{c}{(8)}                     & 
\multicolumn{1}{c}{(9)}                     & 
\multicolumn{1}{c}{(10)}                    & 
\multicolumn{1}{c}{(11)}                    \\ 
% ***********************************************
\noalign{\smallskip}
\hline
\noalign{\smallskip}
\endhead
% ***********************************************
\noalign{\smallskip}
\hline
\endfoot
\noalign{\smallskip}
\hline
\noalign{\smallskip}
% ***********************************************
\multicolumn{11}{p{1.0\linewidth}}{References ---
 (1) \citealt{McConnachie:etal:05}
 (2) \citealt{Rizzi:etal:07}
 (3) \citealt{Karachentsev:etal:06}
 (4) \citealt{Seth:etal:05}
 (5) \citealt{Tully:etal:06}
 (6) \citealt{Karachentsev:etal:04}
 (7) \citealt{Sakai:etal:99}
 (8) \citealt{Tikhonov:06}
 (9) \citealt{Mouhcine:etal:05}
(10) \citealt{Armandroff:etal:99}
(11) \citealt{Macri:etal:06}
(12) \citealt{Tikhonov:Galazutdinova:05}
(13) \citealt{Rizzi:etal:07a}
(14) \citealt{Alonso-Garcia:etal:06}
(15) \citealt{Aloisi:etal:07}
(16) \citealt{Karachentsev:Kashibadze:06}
(17) \citealt{Sakai:etal:04}
(18) \citealt{Sakai:etal:97}
(19) \citealt{Karachentsev:etal:02a}
(20) \citealt{Karachentsev:etal:07}
(21) \citealt{Bellazzini:etal:05}
(22) \citealt{Saviane:etal:04}
(23) \citealt{Tikhonov:etal:06}
(24) \citealt{Rejkuba:etal:05}; \citealt{Karataeva:etal:06};
(25) \citealt{Zucker:etal:06}
(26) \citealt{Aloisi:etal:05}
(27) \citealt{Bellazzini:etal:02}
}
% ***********************************************
\endlastfoot
% ***********************************************
WLM             & LG    & -122 &  -11 &       & 24.87 & 24.82 &       & 24.84 & 24.47 & 1,2     \\
E349-031        &       & 221  &  222 &       & 27.53 &       &       & 27.53 & 27.47 & 3       \\
N0055           & Scl1  & 129  &  117 &       & 26.64 & 26.41 &       & 26.53 & 26.51 & 4,5     \\
E410-05         &       &      &      &       & 26.43 &       &       & 26.43 & 26.34 & 5,6     \\
I0010           & LG    & -348 &  -50 &       & 23.56 &       &       & 23.56 & 21.15 & 7       \\
Sc22            & Scl2  &      &      &       & 28.12 &       &       & 28.12 & 28.02 & 6       \\
Cetus           & LG    &      &      &       & 24.42 &       &       & 24.42 & 23.93 & 1,6     \\
E294-10         &       & 117  &   89 &       & 26.49 &       &       & 26.49 & 26.50 & 6,8     \\
N0147           & LG    & -193 &  103 & 24.20 & 24.27 &       &       & 24.23 & 21.28 & 1,6     \\
And III         & M31   & -351 &  -71 & 24.36 & 24.39 &       &       & 24.38 & 21.70 & 1,6     \\
N0185           & LG    & -202 &   92 & 24.13 & 24.03 &       &       & 24.08 & 20.67 & 2,9     \\
N0205           & M31   & -241 &   48 & 24.65 & 24.59 &       &       & 24.62 & 22.38 & 1,6     \\
And IV          & M31   & 256  &  545 &       & 28.93 &       &       & 28.93 & 28.73 & 6       \\
N0221           & M31   & -200 &   87 &       & 24.43 &       &       & 24.43 & 21.80 & 6       \\
N0224           & M31   & -300 &  -13 & 24.60 & 24.46 & 24.27 &       & 24.44 & 21.83 & 1,2     \\
I1574           &       & 363  &  393 &       & 28.56 &       &       & 28.56 & 28.47 & 6,8     \\
And I           & M31   & -368 &  -87 & 24.44 & 24.44 &       &       & 24.44 & 21.86 & 1,6     \\
N0247           & Scl2  & 156  &  202 &       & 27.81 &       &       & 27.81 & 27.68 & 3       \\
N0253           & Scl2  & 243  &  267 &       & 27.98 &       &       & 27.98 & 27.88 & 6       \\
E540-30         & Scl2  &      &      &       & 27.66 &       &       & 27.66 & 27.50 & 6       \\
E540-31         & Scl2  & 295  &  344 &       & 27.62 &       &       & 27.62 & 27.48 & 6       \\
E540-32         & Scl2  &      &      &       & 27.67 &       &       & 27.67 & 27.52 & 6       \\
SMC             & LG    & 158  &  -24 & 18.98 & 19.00 &       &       & 18.99 & 23.77 & 2       \\
And IX          &       & -216 &   72 &       & 24.40 &       &       & 24.40 & 21.72 & 1       \\
N0300           & Scl1  & 144  &  128 &       & 26.56 & 26.48 &       & 26.52 & 26.49 & 2       \\
Sculptor        & LG    & 110  &  111 & 19.59 & 19.61 &       &       & 19.60 & 23.60 & 2       \\
LGS-3           &       & -287 &  -70 &       & 24.20 &       &       & 24.20 & 22.08 & 1,6     \\
I1613           & LG    & -234 &  -65 & 24.35 & 24.33 & 24.32 &       & 24.33 & 23.35 & 2       \\
U685            &       & 157  &  353 &       & 28.38 &       &       & 28.38 & 28.15 & 5,6     \\
KKH5            &       & 61   &  368 &       & 28.15 &       &       & 28.15 & 27.86 & 6       \\
N0404           &       & -48  &  221 &       & 27.43 &       &       & 27.43 & 27.01 & 6       \\
And V           & M31   & -403 & -121 &       & 24.47 &       &       & 24.47 & 22.07 & 1,10    \\
And II          & M31   & -188 &   90 & 24.15 & 24.11 &       &       & 24.13 & 21.14 & 1,6     \\
UA17            & Cet   & 1959 & 1940 &       &       &       & 33.18 & 33.18 & 33.16 &         \\
N0598           & LG    & -179 &   70 & 24.77 & 24.66 & 24.64 &       & 24.69 & 22.85 & 1,2     \\
KKH6            &       & 53   &  352 &       & 27.86 &       &       & 27.86 & 27.53 & 3       \\
N0625           &       & 396  &  338 &       & 28.05 &       &       & 28.05 & 28.04 & 6       \\
E245-05         &       & 391  &  319 &       & 28.23 &       &       & 28.23 & 28.23 & 6       \\
U1281           &       & 156  &  399 &       & 28.55 &       &       & 28.55 & 28.32 & 5,8     \\
Phoenix         & LG    & 56   &  -16 & 23.05:& 23.22 &       &       & 23.22 & 24.16 & 6       \\
KK16            &       & 207  &  430 &       & 28.62 &       &       & 28.62 & 28.40 & 5,11    \\
KK17            &       & 168  &  394 &       & 28.41 &       &       & 28.41 & 28.17 & 5,6     \\
N0784           &       & 198  &  423 &       & 28.58 &       &       & 28.58 & 28.36 & 5       \\
N0891           &       & 528  &  793 &       & 29.96 &       &       & 29.96 & 29.84 & 12      \\
N0925           &       & 553  &  782 &       &       & 29.84 &       & 29.84 & 29.72 &         \\
E115-21         &       & 515  &  373 &       & 28.43 &       &       & 28.43 & 28.50 & 5,8     \\
Fornax          & LG    & 53   &    3 & 20.67 & 20.72 &       &       & 20.70 & 23.64 & 13      \\
E154-23         &       & 574  &  444 &       & 28.80 &       &       & 28.80 & 28.84 & 5       \\
KKH18           &       & 216  &  437 &       & 28.23 &       &       & 28.23 & 27.99 & 6       \\
N1313           &       & 470  &  307 &       & 28.15 &       &       & 28.15 & 28.26 & 2       \\
N1311           &       & 568  &  439 &       & 28.68 &       &       & 28.68 & 28.73 & 5       \\
KK27            &       &      &      &       & 28.04 &       &       & 28.04 & 28.16 & 5,6     \\
N1316           & For   & 1760 & 1371 &       &       &       & 31.48 & 31.48 & 31.48 &         \\
N1326A          & For   & 1831 & 1371 &       &       & 31.17 &       & 31.17 & 31.17 &         \\
I1959           &       & 640  &  511 &       & 28.91 &       &       & 28.91 & 28.95 & 5       \\
N1365           & For   & 1636 & 1371 &       &       & 31.46 &       & 31.46 & 31.46 &         \\
N1380           & For   & 1877 & 1371 &       &       &       & 31.81 & 31.81 & 31.81 &         \\
N1425           & For   & 1510 & 1371 &       &       & 31.96 &       & 31.96 & 31.95 &         \\
N1448           &       & 1168 & 1015 &       &       &       & 31.78 & 31.78 & 31.79 &         \\
KK35            & I342  & 105  &  382 &       & 27.50 &       &       & 27.50 & 27.19 & 6       \\
UA86            & I342  & 67   &  337 &       & 27.36 &       &       & 27.36 & 27.04 & 3       \\
Cam A           & I342  & -46  &  232 &       & 27.97 &       &       & 27.97 & 27.74 & 6       \\
UA92            & I342  & -99  &  155 &       & 27.39 &       &       & 27.39 & 27.09 & 3       \\
N1560           & I342  & -36  &  234 &       & 27.70 &       &       & 27.70 & 27.44 & 6,8     \\
N1637           &       & 717  &  740 &       &       & 30.40 &       & 30.40 & 30.37 &         \\
Cam B           & I342  & 77   &  335 &       & 27.62 &       &       & 27.62 & 27.36 & 6       \\
N1705           &       & 633  &  474 &       & 28.54 &       &       & 28.54 & 28.62 & 6       \\
UA105           & I342  & 111  &  351 &       & 27.49 &       &       & 27.49 & 27.23 & 6       \\
LMC             & LG    & 278  &   42 & 18.53 & 18.59 &       &       & 18.56 & 23.78 & 2       \\
N2090           &       & 921  &  810 &       &       & 30.48 &       & 30.48 & 30.50 &         \\
KKH34           &       & 110  &  374 &       & 28.32 &       &       & 28.32 & 28.15 & 6       \\
E121-20         &       & 575  &  390 &       & 28.91 &       &       & 28.91 & 29.01 & 3       \\
E489-56         &       & 492  &  371 &       & 28.49 &       &       & 28.49 & 28.56 & 6       \\
E490-17         &       & 504  &  371 &       & 28.13 &       &       & 28.13 & 28.22 & 6       \\
Carina          & LG    & 229  &  -14 & 20.09 & 20.00 &       &       & 20.05 & 23.89 & 6       \\
KKH37           &       & -148 &  106 &       & 27.65 &       &       & 27.65 & 27.43 & 3       \\
FG202           &       & 564  &  358 &       & 28.45 &       &       & 28.45 & 28.60 & 6       \\
U3755           &       & 315  &  335 &       & 29.35 &       &       & 29.35 & 29.35 & 5,11    \\
DDO43           &       & 354  &  507 &       & 29.46 &       &       & 29.46 & 29.42 & 6       \\
N2366           & N2403 & 80   &  293 &       & 27.55 &       &       & 27.55 & 27.36 & 11      \\
E059-01         &       & 530  &  312 &       & 28.30 &       &       & 28.30 & 28.47 & 3       \\
DDO44           & N2403 &      &      &       & 27.52 &       &       & 27.52 & 27.34 & 6,14    \\
N2403           & N2403 & 131  &  327 &       &       & 27.43 &       & 27.43 & 27.25 &         \\
DDO47           &       & 272  &  309 &       & 29.53 &       &       & 29.53 & 29.53 & 5       \\
KK65            &       & 279  &  314 &       & 29.52 &       &       & 29.52 & 29.52 & 5       \\
U4115           &       & 341  &  352 &       & 29.44 &       &       & 29.44 & 29.46 & 5       \\
N2541           &       & 548  &  780 &       &       & 30.50 &       & 30.50 & 30.47 &         \\
Ho II           & N2403 & 142  &  350 &       & 27.65 &       &       & 27.65 & 27.49 & 6       \\
KDG52           & N2403 & 113  &  322 &       & 27.75 &       &       & 27.75 & 27.59 & 6       \\
DDO52           &       & 397  &  555 &       & 30.06 &       &       & 30.06 & 30.04 & 3       \\
DDO53           & N2403 & 20   &  204 &       & 27.76 &       &       & 27.76 & 27.63 & 6       \\
U4483           & N2403 & 156  &  354 &       & 27.53 &       &       & 27.53 & 27.37 & 6       \\
D564-08         &       & 483  &  473 &       & 29.69 &       &       & 29.69 & 29.72 & 3       \\
D634-03         &       & 318  &  290 &       & 29.90 &       &       & 29.90 & 29.94 & 3       \\
D565-06         &       & 498  &  483 &       & 29.79 &       &       & 29.79 & 29.82 & 3       \\
N2841           &       & 638  &  882 &       &       & 30.75 &       & 30.75 & 30.73 &         \\
U4998           &       & 623  &  870 &       & 29.63 &       &       & 29.63 & 29.57 & 14      \\
N2915           &       & 468  &  238 &       & 27.89 &       &       & 27.89 & 28.12 & 6       \\
I Zw 18         &       & 751  &  971 &       & 30.32 &       &       & 30.32 & 30.30 & 15      \\
Ho I            & M81   & 139  &  337 &       & 27.92 &       &       & 27.92 & 27.80 & 6       \\
F8D1            & M81   &      &      &       & 27.88 &       &       & 27.88 & 27.77 & 6       \\
FM1             & M81   &      &      &       & 27.67 &       &       & 27.67 & 27.55 & 6       \\
N2976           & M81   & 3    &  179 &       & 27.76 &       &       & 27.76 & 27.64 & 6       \\
KK77            & M81   &      &      &       & 27.71 &       &       & 27.71 & 27.60 & 6       \\
N3021           &       & 1541 & 1840 &       &       &       & 32.62 & 32.62 & 32.62 &         \\
BK3N            & M81   & -40  &  145 &       & 28.02 &       &       & 28.02 & 27.91 & 6       \\
N3031           & M81   & -34  &  147 &       & 27.80 & 27.80 &       & 27.80 & 27.68 & 2       \\
N3034           & M81   & 203  &  390 &       & 27.85 &       &       & 27.85 & 27.73 & 6,8     \\
KDG61           & M81   & -135 &   42 &       & 27.78 &       &       & 27.78 & 27.67 & 6       \\
Ho IX           & M81   & 46   &  228 &       & 27.84 &       &       & 27.84 & 27.73 & 16      \\
A0952+69        & M81   & 99   &  285 &       & 27.94 &       &       & 27.94 & 27.83 & 6       \\
Leo A           & LG    & 24   &  -12 & 24.54 & 24.19 &       &       & 24.37 & 24.97 & 6       \\
SexB            & LG    & 300  &  138 &       & 25.75 &       &       & 25.75 & 26.21 & 2       \\
KKH57           & M81   &      &      &       & 27.97 &       &       & 27.97 & 27.89 & 6       \\
N3109           & LG    & 403  &  129 &       & 25.54 & 25.45 &       & 25.50 & 26.18 & 2       \\
N3077           & M81   & 14   &  194 &       & 27.91 &       &       & 27.91 & 27.80 & 6       \\
Antlia          & LG    & 362  &   85 &       & 25.55 &       &       & 25.55 & 26.22 & 5,6     \\
BK5N            & M81   &      &      &       & 27.89 &       &       & 27.89 & 27.78 & 6       \\
KDG63           & M81   & -129 &   34 &       & 27.72 &       &       & 27.72 & 27.62 & 6       \\
KDG64           & M81   & -18  &  155 &       & 27.84 &       &       & 27.84 & 27.73 & 6       \\
U5456           &       & 544  &  391 &       & 27.90 &       &       & 27.90 & 28.05 & 6       \\
IKN             & M81   &      &      &       & 27.87 &       &       & 27.87 & 27.76 & 3       \\
Leo I           & LG    & 285  &  154 & 22.01 &       &       &       & 22.01 & 24.19 &         \\
SexA            & LG    & 324  &  117 &       & 25.74 &       &       & 25.74 & 26.28 & 2       \\
Sex dSph        & LG    & 224  &   29 & 19.69 & 19.77 &       &       & 19.73 & 23.88 & 6       \\
N3190           &       & 1271 & 1574 &       &       &       & 32.15 & 32.15 & 32.16 &         \\
N3198           &       & 663  &  858 &       &       & 30.80 &       & 30.80 & 30.80 &         \\
HS117           & M81   & -37  &  155 &       & 27.99 &       &       & 27.99 & 27.88 & 3       \\
DDO78           & M81   & 55   &  226 &       & 27.85 &       &       & 27.85 & 27.75 & 6       \\
I2574           & M81   & 57   &  235 &       & 28.02 &       &       & 28.02 & 27.92 & 6       \\
DDO82           & M81   & 56   &  246 &       & 28.01 &       &       & 28.01 & 27.90 & 6       \\
BK6N            & M81   &      &      &       & 27.93 &       &       & 27.93 & 27.84 & 6       \\
N3319           &       & 739  &  878 &       &       & 30.74 &       & 30.74 & 30.75 &         \\
N3351           & LeoI  & 778  &  588 &       & 30.23 & 30.10 &       & 30.17 & 30.23 & 2,17    \\
N3368           & LeoI  & 897  &  715 &       &       & 30.34 & 30.50 & 30.42 & 30.47 &         \\
N3370           &       & 1279 & 1606 &       &       & 32.37 & 32.47 & 32.42 & 32.44 &         \\
N3379           & LeoI  & 911  &  721 &       & 30.32 &       &       & 30.32 & 30.37 & 18      \\
KDG73           &       & 116  &  297 &       & 27.91 &       &       & 27.91 & 27.81 & 19      \\
E215-09         &       & 598  &  345 &       & 28.60 &       &       & 28.60 & 28.80 & 20      \\
Leo II          & LG    & -87  & -172 & 21.58 & 21.72 &       &       & 21.65 & 24.08 & 6,21    \\
N3621           &       & 730  &  487 &       & 29.27 & 29.30 &       & 29.29 & 29.44 & 2       \\
N3627           & LeoI  & 727  &  428 &       &       & 30.50 & 30.41 & 30.46 & 30.51 &         \\
U6456           &       & -103 &  133 &       & 28.19 &       &       & 28.19 & 28.06 & 6,8     \\
U6541           & CVn   & 250  &  297 &       & 27.95 &       &       & 27.95 & 27.96 & 6       \\
N3738           & CVn   & 229  &  316 &       & 28.45 &       &       & 28.45 & 28.43 & 6       \\
N3741           & CVn   & 229  &  251 &       & 27.46 &       &       & 27.46 & 27.51 & 5,6     \\
E320-14         &       & 654  &  402 &       & 28.92 &       &       & 28.92 & 29.10 & 20      \\
KK109           & CVn   & 212  &  217 &       & 28.27 &       &       & 28.27 & 28.30 & 6       \\
DDO99           &       & 242  &  228 &       & 27.11 &       &       & 27.11 & 27.22 & 5,6     \\
E379-07         &       & 641  &  376 &       & 28.59 &       &       & 28.59 & 28.80 & 6       \\
N3982           & UMa   & 1109 & 1515 &       &       & 31.87 & 32.02 & 31.94 & 31.93 &         \\
N4038           &       & 1642 & 1435 &       & 30.46 &       &       & 30.46 & 30.55 & 22      \\
N4068           &       & 210  &  282 &       & 28.17 &       &       & 28.17 & 28.17 & 3       \\
N4144           &       & 265  &  294 &       & 29.32 &       &       & 29.32 & 29.33 & 4,12    \\
N4163           &       & 165  &  132 &       & 27.35 &       &       & 27.35 & 27.46 & 3,5     \\
E321-14         &       & 610  &  335 &       & 27.52 &       &       & 27.52 & 27.86 & 6,8     \\
U7242           & N4236 & 68   &  243 &       & 28.67 &       &       & 28.67 & 28.61 & 3       \\
DDO113          &       & 284  &  253 &       & 27.40 &       &       & 27.40 & 27.51 & 5,6     \\
N4214           &       & 291  &  262 &       & 27.34 &       &       & 27.34 & 27.45 & 5,6     \\
U7298           & CVn   & 173  &  243 &       & 28.12 &       &       & 28.12 & 28.12 & 6       \\
N4236           & N4236 & 0    &  187 &       & 28.24 &       &       & 28.24 & 28.16 & 6       \\
N4244           & CVn   & 244  &  212 &       & 28.09 &       &       & 28.09 & 28.16 & 4,9,12  \\
I3104           &       & 429  &  191 &       & 26.80 &       &       & 26.80 & 27.18 & 6,8     \\
N4258           &       & 448  &  488 &       & 29.32 & 29.50 &       & 29.41 & 29.42 & 9,11    \\
I0779           &       & 222  &    7 &       & 30.32 &       &       & 30.32 & 30.36 & 3       \\
N4321           & Vir A & 1571 & 1152 &       &       & 31.18 &       & 31.18 & 31.22 &         \\
N4395           & CVn   & 319  &  258 &       & 28.32 & 28.02 &       & 28.17 & 28.25 & 6       \\
N4414           &       & 716  &  983 &       &       & 31.65 & 31.28 & 31.46 & 31.48 &         \\
N4419           & Vir A & -261 & 1152 &       &       &       & 31.15 & 31.15 & 31.19 &         \\
DDO126          & CVn   & 218  &  176 &       & 28.44 &       &       & 28.44 & 28.50 & 6       \\
DDO125          &       & 195  &  215 &       & 27.11 &       &       & 27.11 & 27.19 & 5,6     \\
N4449           & CVn   & 207  &  221 &       & 28.12 &       &       & 28.12 & 28.16 & 6       \\
U7605           & CVn   & 310  &  263 &       & 28.23 &       &       & 28.23 & 28.30 & 6       \\
N4496A          & Vir W & 1730 & 1075 &       &       & 31.18 & 30.77 & 30.97 & 31.02 &         \\
N4501           & Vir A & 2281 & 1152 &       &       &       & (30.84) & \nodata & \nodata &   \\
N4526           & Vir B & 448  & 1152 &       &       &       & 31.30 & 31.30 & 31.34 &         \\
N4527           & Vir W & 1736 & 1204 &       &       & 30.76 &       & 30.76 & 30.82 &         \\
N4535           & Vir B & 1961 & 1152 &       &       & 31.25 &       & 31.25 & 31.29 &         \\
N4536           & Vir W & 1808 & 1424 &       &       & 31.24 & 31.28 & 31.26 & 31.31 &         \\
N4548           & Vir A & 486  & 1152 &       &       & 30.99 &       & 30.99 & 31.03 &         \\
Arp211          &       & 458  &  419 &       & 29.13 &       &       & 29.13 & 29.17 & 6       \\
N4605           &       & 143  &  292 &       & 28.72 &       &       & 28.72 & 28.68 & 2       \\
N4631           &       & 606  &  501 &       & 29.42 &       &       & 29.42 & 29.47 & 4       \\
I3687           & CVn   & 354  &  330 &       & 28.30 &       &       & 28.30 & 28.36 & 6       \\
N4639           & Vir A & 1018 & 1152 &       &       & 32.20 & 32.05 & 32.12 & 32.15 &         \\
E381-18         &       & 624  &  371 &       & 28.55 &       &       & 28.55 & 28.77 & 8,20    \\
E381-20         &       & 589  &  338 &       & 28.68 &       &       & 28.68 & 28.88 & 20      \\
HI J1247-77     &       & 413  &  181 &       & 27.50 &       &       & 27.50 & 27.79 & 3       \\
KK166           & CVn   &      &      &       & 28.38 &       &       & 28.38 & 28.45 & 6       \\
N4725           &       & 1206 &  904 &       &       & 30.65 &       & 30.65 & 30.69 &         \\
N4736           & CVn   & 308  &  306 &       & 28.34 &       &       & 28.34 & 28.39 & 6       \\
N4753           &       & 1239 & 1310 &       &       &       & 31.41 & 31.41 & 31.46 &         \\
E443-09         &       & 645  &  397 &       & 28.88 &       &       & 28.88 & 29.06 & 20      \\
DDO155          &       & 214  &   88 &       & 26.63 &       &       & 26.63 & 26.96 & 5,6     \\
E269-37         & CenA  &      &      &       & 27.71 &       &       & 27.71 & 28.02 & 6       \\
KK182           &       & 617  &  381 &       & 28.81 &       &       & 28.81 & 29.00 & 20      \\
N4945           & CenA  & 563  &  300 &       & 27.25 &       &       & 27.25 & 27.63 & 9       \\
I4182           & CVn   & 321  &  301 &       & 28.19 & 28.21 & 28.45 & 28.28 & 28.34 & 2       \\
DDO165          &       & 31   &  216 &       & 28.30 &       &       & 28.30 & 28.23 & 6       \\
U8215           & N4236 & 218  &  264 &       & 28.29 &       &       & 28.29 & 28.31 & 3       \\
E269-58         & CenA  & 400  &  148 &       & 27.90 &       &       & 27.90 & 28.19 & 20      \\
N5023           &       & 407  &  433 &       & 29.02 &       &       & 29.02 & 29.04 & 4,23    \\
KK189           & CenA  &      &      &       & 28.23 &       &       & 28.23 & 28.48 & 20      \\
E269-66         & CenA  & 784  &  533 &       & 27.91 &       &       & 27.91 & 28.20 & 20      \\
DDO167          & CVn   & 163  &  208 &       & 28.11 &       &       & 28.11 & 28.14 & 6       \\
DDO168          & CVn   & 192  &  235 &       & 28.18 &       &       & 28.18 & 28.21 & 6       \\
KK195           & M83   & 571  &  334 &       & 28.59 &       &       & 28.59 & 28.80 & 6       \\
KK196           & CenA  & 741  &  495 &       & 28.00 &       &       & 28.00 & 28.27 & 20      \\
N5102           & CenA  & 468  &  218 &       & 27.66 &       &       & 27.66 & 27.98 & 6       \\
KK197           & CenA  &      &      &       & 27.94 &       &       & 27.94 & 28.22 & 20      \\
KKs55           & CenA  &      &      &       & 27.98 &       &       & 27.98 & 28.26 & 20      \\
KK200           & M83   & 487  &  248 &       & 28.33 &       &       & 28.33 & 28.56 & 6       \\
N5128           & CenA  & 547  &  298 &       & 27.89 & 27.67 &       & 27.78 & 28.08 & 6,24    \\
I4247           & M83   & 274  &   38 &       & 28.48 &       &       & 28.48 & 28.70 & 20      \\
E324-24         & CenA  & 516  &  270 &       & 27.86 &       &       & 27.86 & 28.15 & 6       \\
CVn dSph        & LG    & 36   &   46 &       & 21.83 &       &       & 21.83 & 24.03 & 25      \\
N5204           & CVn   & 201  &  336 &       & 28.34 &       &       & 28.34 & 28.31 & 6       \\
U8508           &       & 62   &  169 &       & 27.10 &       &       & 27.10 & 27.09 & 5,6     \\
N5206           & CenA  & 571  &  325 &       & 27.70 &       &       & 27.70 & 28.01 & 20      \\
E444-78         & M83   & 573  &  346 &       & 28.60 &       &       & 28.60 & 28.81 & 20      \\
KK208           & M83   & 381  &  150 &       & 28.35 &       &       & 28.35 & 28.58 & 6       \\
DE J1337-33     & M83   & 591  &  358 &       & 28.27 &       &       & 28.27 & 28.51 & 6       \\
N5236           & M83   & 513  &  283 &       & 28.56 & 28.32 &       & 28.44 & 28.66 & 20      \\
E444-084        & CenA  & 587  &  357 &       & 28.32 &       &       & 28.32 & 28.55 & 6       \\
HI J1337-39     &       & 492  &  262 &       & 28.45 &       &       & 28.45 & 28.67 & 6       \\
N5237           & CenA  & 361  &  116 &       & 27.66 &       &       & 27.66 & 27.98 & 20      \\
U8638           &       & 274  &  198 &       & 28.15 &       &       & 28.15 & 28.27 & 3       \\
DDO181          &       & 202  &  231 &       & 27.40 &       &       & 27.40 & 27.48 & 5,6     \\
N5253           & CenA  & 407  &  172 &       & 27.89 & 28.05 & 27.95 & 27.96 & 28.23 & 17      \\
I4316           & M83   & 674  &  444 &       & 28.22 &       &       & 28.22 & 28.46 & 6       \\
N5264           & M83   & 478  &  249 &       & 28.28 &       &       & 28.28 & 28.52 & 6       \\
KKs57           & CenA  &      &      &       & 27.97 &       &       & 27.97 & 28.25 & 20      \\
KK211           & CenA  &      &      &       & 27.77 &       &       & 27.77 & 28.07 & 6       \\
KK213           & CenA  &      &      &       & 27.80 &       &       & 27.80 & 28.10 & 6       \\
E325-11         & CenA  & 545  &  304 &       & 27.66 &       &       & 27.66 & 27.97 & 6       \\
KK217           & CenA  &      &      &       & 27.92 &       &       & 27.92 & 28.20 & 6       \\
CenN            & CenA  &      &      &       & 27.88 &       &       & 27.88 & 28.16 & 20      \\
KK221           & CenA  &      &      &       & 28.00 &       &       & 28.00 & 28.27 & 6       \\
HI 1348-37      &       & 581  &  367 &       & 28.80 &       &       & 28.80 & 28.99 & 20      \\
E383-87         & CenA  & 326  &   91 &       & 27.69 &       &       & 27.69 & 28.00 & 20      \\
DDO183          &       & 192  &  211 &       & 27.55 &       &       & 27.55 & 27.63 & 5       \\
HI 1351-47      &       & 529  &  317 &       & 28.79 &       &       & 28.79 & 28.98 & 20      \\
KKH86           &       & 287  &  148 &       & 27.08 &       &       & 27.08 & 27.38 & 5,6     \\
U8833           & CVn   & 227  &  236 &       & 27.52 &       &       & 27.52 & 27.62 & 5,6     \\
E384-016        & CenA  & 561  &  340 &       & 28.28 &       &       & 28.28 & 28.52 & 20      \\
N5457           &       & 241  &  387 &       & 29.39 & 29.17 &       & 29.28 & 29.27 & 2,17    \\
N5408           &       & 506  &  289 &       & 28.41 &       &       & 28.41 & 28.63 & 6       \\
KK230           &       & 62   &   82 &       & 26.54 &       &       & 26.54 & 26.71 & 3,5     \\
DDO187          &       & 153  &  117 &       & 26.87 &       &       & 26.87 & 27.09 & 5,6     \\
SBS1415+437     &       & 609  &  805 &       & 30.70 &       &       & 30.70 & 30.71 & 26      \\
DDO190          &       & 150  &  229 &       & 27.23 &       &       & 27.23 & 27.28 & 5,6     \\
P51659          & CenA  & 390  &  172 &       & 27.77 &       &       & 27.77 & 28.06 & 6       \\
E223-09         &       & 588  &  423 &       & 29.06 &       &       & 29.06 & 29.22 & 20      \\
UMi             & LG    & -247 &  -57 & 19.29 & 19.51 &       &       & 19.40 & 23.56 & 27      \\
E274-01         &       & 522  &  325 &       & 27.45 &       &       & 27.45 & 27.77 & 20      \\
KKR25           &       & -139 &   44 &       & 26.50 &       &       & 26.50 & 26.42 & 5,6     \\
E137-18         &       & 605  &  456 &       & 29.03 &       &       & 29.03 & 29.17 & 20      \\
Draco           & LG    & -292 &  -75 & 19.59 & 19.92 &       &       & 19.76 & 23.53 & 27      \\
I4662           &       & 302  &  135 &       & 26.94 &       &       & 26.94 & 27.26 & 3       \\
N6503           &       & 60   &  357 &       & 28.61 &       &       & 28.61 & 28.49 & 6       \\
Sag dSph        & LG    & 140  &  101 & 17.22 & 16.51 &       &       & 16.87 & 23.69 & 6       \\
N6789           &       & -141 &  162 &       & 27.78 &       &       & 27.78 & 27.58 & 6       \\
Sag DIG         & LG    & -79  &  -37 &       & 25.09 &       &       & 25.09 & 25.39 & 6       \\
N6822           & LG    & -57  &    7 & 23.43 & 23.37 & 23.31 &       & 23.37 & 24.25 & 17      \\
E461-36         &       & 427  &  454 &       & 29.47 &       &       & 29.47 & 29.49 & 3       \\
N6951           &       & 1424 & 1814 &       &       &       & 31.89 & 31.89 & 31.85 &         \\
DDO210          & LG    & -141 &  -36 &       & 25.01 &       &       & 25.01 & 25.05 & 1,6     \\
I5052           &       & 584  &  455 &       & 28.89 &       &       & 28.89 & 28.99 & 4       \\
I5152           &       & 122  &   63 &       & 26.52 &       &       & 26.52 & 26.68 & 5,6     \\
N7331           &       & 816  & 1099 &       &       & 30.89 &       & 30.89 & 30.82 &         \\
Tucana          & LG    & 130  &   -6 &       & 24.72 &       &       & 24.72 & 25.34 & 6       \\
I5270           &       & 1983 & 1914 &       &       &       & 31.90 & 31.90 & 31.89 &         \\
UA438           &       & 62   &   89 &       & 26.74 &       &       & 26.74 & 26.67 & 5,6     \\
Cas dSph        & LG    & -307 &    0 &       & 24.45 &       &       & 24.45 & 22.37 & 1,6     \\
Pegasus         & LG    & -183 &   61 &       & 24.60 &       &       & 24.60 & 23.32 & 1,6     \\
UA442           &       & 267  &  276 &       & 28.24 &       &       & 28.24 & 28.18 & 6,8     \\
KKH98           &       & -137 &  162 &       & 26.95 &       &       & 26.95 & 26.43 & 6       \\
And VI          & M31   & -354 & -103 & 24.59 & 24.48 &       &       & 24.53 & 22.71 & 1,10    \\
N7793           &       & 227  &  234 &       & 27.96 &       &       & 27.96 & 27.90 & 6       \\
% ***********************************************
\end{longtable}
\normalsize
% ***********************************************

% ********************************************************
% Figure 1: 4 Hubble diagrams
% ********************************************************
% ********************************************************

% ********************************************************
% 2.3 The Hubble diagram of Cepheid distances
% ********************************************************
\subsection{The Hubble diagram of Cepheid distances}
\label{sec:2:3}
The 37 Cepheid distances in Table~\ref{tab:dist} are plotted in a Hubble
diagram in Fig.~\ref{fig:hubble}b. A linear regression,
omitting seven galaxies with $\mu^{00}<28.2$ and the deviating case of
NGC\,3627, gives a free orthogonal fit for the slope of $0.200\pm0.010$
in excellent agreement with linear expansion. 

     The dispersion about the Hubble line is small at
$0.34\mag$. Subtracting in quadrature $0.15\mag$ for random errors of
the Cepheid moduli leaves a contribution of $0.30\mag$ for the
peculiar velocities.

% ********************************************************
% 2.4 The Hubble diagram of SNe Ia
% ********************************************************
\subsection{The Hubble diagram of SNe\,Ia}
\label{sec:2:4}
22 SNe\,Ia distances are listed in Table~\ref{tab:dist}. 
Omitting SN\,1937C in IC\,4182, which has $\mu^{00}<28.2$, and the
deviating SN\,1989B in NGC\,3627 yields an orthogonal fit for the
Hubble line with slope $0.192\pm0.016$, giving additional support for
the nearly  perfect linear expansion with slope $0.2$
(Fig.~\ref{fig:hubble}c). The dispersion is $\sigma_{m}=0.43$ in $B$,
$V$, and $I$.

     In addition there are 62 SNe\,Ia with $3000<v_{220}<20,000\kms$
\citep[Fig. 15 in][]{RTS:05} whose magnitudes are uniformly reduced as in
the case of the nearer SNe\,Ia. They give an orthogonal slope of
$0.194\pm0.002$ which is significantly smaller than 0.2, but it is
almost exactly the value predicted for a linearly expanding flat
Universe with $\Omega_{\Lambda}=0.7$ \citep{Carroll:etal:92}.

     The scatter about the Hubble line in $B$, $V$, and $I$ beyond 
$v_{\rm CMB}=3000\kms$ is only $\sigma=0.14\mag$ after absorption
corrections and normalization to a fiducial decline rate; in dust-poor
S0 and E galaxies it is even smaller. The small scatter is a
confirmation that properly reduced SNe\,Ia yield distance moduli to
within $0.15\mag$ as claimed above. Differently treated SNe\,Ia by
\citet{Wang:etal:06} lead essentially to the same results.  

\citet{Wood-Vasey:etal:07} have constructed a Hubble diagram from the
near-infrared $H$ magnitudes, which are less affected by absorption,
of 32 SNe\,Ia in the distance range $2000<v_{220}<10,000\kms$. Again
the slope is as close to 0.2 as can be measured. The scatter amounts
to only $0.15\mag$ even without normalization to a fixed decline rate
or light curve width. 

     \citet{Jha:etal:07} have presented a Hubble diagram with a
dispersion of $\sigma_{m}=0.18$ for 95 SNe\,Ia with 
$2500<v_{\rm CMB}<40,000\kms$. At low redshifts its asymptotic slope
is very close to $0.2$ and fits at higher redshifts the slope
corresponding to $\Omega_{M}=0.3$, $\Omega_{\Lambda}=0.7$. 
Yet the authors, reviving similar suggestions by \citet{Tammann:98}
and \citet{Zehavi:etal:98}, propose a break of the Hubble line of
SNe\,Ia at $\sim\!7400\kms$, implying a decrease of $H_{0}$ at larger
distances by $\sim\!6.5\%$, but the effect is not seen in the
aforementioned studies. 

     There are other relative distance indicators which confirm the
linearity of the expansion field. They are not on a uniform zero
point, but strengthen the conclusion of linearity or are at least in
agreement with it. The difficulty is in general the large intrinsic
scatter which prohibits a stringent test. A way out is to use mean
cluster distances from a subset of cluster members. Examples of 
{\em relative\/} cluster distances reaching out to 
$\sim\!10,000\kms$ are in \citet{Dressler:87},
\citet{Lynden-Bell:etal:88}, and \citet{Jerjen:Tammann:93}. 
The mean distances of 10 clusters with about 20 $D_{n}\!-\!\sigma$ 
distances each are given by \citet[][see also \citealt{Tammann:06},
Fig.~7]{Jorgensen:etal:96}. 
\citet{Hudson:etal:04} have derived relative distances of 56 Abell
clusters within $12,000\kms$ from an inverse fit to the fundamental
plane relation (FP); they find local streaming motions, but their
overall expansion is linear in close approximation.

     Also the mean distances of 31 clusters with about 15 21cm line
width (TF) distances each  \citep{Masters:etal:06} define a Hubble
line for $1000<v_{\rm CMB}<10,000\kms$ with a dispersion of
$0.12\mag$.  
The latter sample illustrates the inherent problem to select a 
{\em fair\/} subset of cluster members independent of distance: 
their three {\em nearest\/} clusters fall systematically off the 
Hubble line (\citeauthor*{TSR:08}, Fig.~8), whose slope is otherwise 
almost precisely 0.2.

% ********************************************************
% 2.5 Characteristics of the expansion field
% ********************************************************
\subsection{Characteristics of the expansion field}
\label{sec:2:5}
The evidence from relative TRGB, Cepheid, and SNe\,Ia distances in 
Sect.~\ref{sec:2:2}-\ref{sec:2:4} strongly confines the
all-sky-averaged deviations from linear expansion and shows that a
{\em single value of\/}$H_{0}$ applies for all practical purposes from 
$\sim 250<v_{220}<20,000$ or even $30,000\kms$, 
at which distance the cosmic value of $H_{0}$ must be reached for all
classical models. Moreover, the dispersion about the Hubble line is in
some cases significantly larger than the observational error of the
distance indicators. In these cases it is possible to give meaningful
estimates of the random motion of field galaxies. The results are laid
out in Table~1. In Col.~1 the distance range (in Mpc nearby and in $\kms$
for the more distant galaxies) is given for a particular distance
indicator in Col.~2 with the number of galaxies involved in Col.~3. 
The free-fit slope of the Hubble line for $\log v$ versus $\mu^{00}$
(or $m^{0}$) is in Col.~4.
The slopes for the inverse and orthogonal regressions are in Cols.~5
and 6, respectively.
The median velocity of the sample follows in Col.~7. The observed
magnitude dispersion is shown in Col.~8 for the case of a fixed slope
of 0.2. The dispersion is reduced in quadrature for the mean
observational error of the distance determination, which is assumed to
be $0.15\mag$ for the distance indicators used. The remaining scatter
must be due to peculiar velocities. Multiplying the magnitude scatter
by 0.2 leads to the scatter in $\log v_{220}$ and hence to $v_{\rm
  pec}/v_{220}$ shown in Col.~9. The product of the latter and the
corresponding median velocity yields an estimate of the mean peculiar
velocity (Col.~10) at the distance of the median velocity.
Finally the intercept $a$ for the case of a forced slope of 0.2 in
Col.~11 and the value of $H_{0}$ in Col.~12 will be discussed in
Sect.~\ref{sec:3}. 

% ***********************************************
%  Table 2: Characteristics of the Expansion Field
% ***********************************************
\begin{sidewaystable}
% \begin{table}
\begin{center}
\caption{ Characteristics of the Expansion Field.} 
\label{tab:char}
%\scriptsize
\tiny
\begin{tabular}{lcccccrrrrcc}
% ***********************************************
\hline
\hline
\noalign{\smallskip}
% ***********************************************
                                            & 
\multicolumn{1}{c}{distance}                & 
                                            & 
\multicolumn{1}{c}{slope}                   & 
\multicolumn{1}{c}{slope}                   & 
\multicolumn{1}{c}{slope}                   & 
\multicolumn{1}{c}{$v_{220}$}               & 
                                            & 
                                            &
                                            & 
\multicolumn{1}{c}{$a$}                     & 
                                            \\
\multicolumn{1}{c}{range}                   & 
\multicolumn{1}{c}{indicator}               & 
\multicolumn{1}{c}{n}                       & 
\multicolumn{1}{c}{direct}                  & 
\multicolumn{1}{c}{inverse}                 & 
\multicolumn{1}{c}{orthogonal}              & 
\multicolumn{1}{c}{median}                  & 
\multicolumn{1}{c}{$\sigma_{m}$}            & 
\multicolumn{1}{c}{$v_{\rm pec}/v_{220}$}   &
\multicolumn{1}{c}{$v_{\rm pec}$}           & 
\multicolumn{1}{c}{(0.2 fixed)}             & 
\multicolumn{1}{c}{$H_{0}$}                 \\
\multicolumn{1}{c}{(1)}                     & 
\multicolumn{1}{c}{(2)}                     & 
\multicolumn{1}{c}{(3)}                     & 
\multicolumn{1}{c}{(4)}                     & 
\multicolumn{1}{c}{(5)}                     & 
\multicolumn{1}{c}{(6)}                     & 
\multicolumn{1}{c}{(7)}                     & 
\multicolumn{1}{c}{(8)}                     & 
\multicolumn{1}{c}{(9)}                     & 
\multicolumn{1}{c}{(10)}                    & 
\multicolumn{1}{c}{(11)}                    & 
\multicolumn{1}{c}{(12)}                    \\
% ***********************************************
\noalign{\smallskip}
\hline
\noalign{\smallskip}
% ***********************************************
3.9-4.4$\;$Mpc & TRGB & 20 & 
\nodata & \nodata & \nodata & 
282 & (0.74) & (0.41) & (114) & 
$-3.180\pm0.034$ & $66.1\pm5.2$ \\ 
$>4.4\;$Mpc & TRGB & 78 & 
$0.166\pm0.019$ & $0.332\pm0.038$ & $0.199\pm0.019$ & 
371 & 0.47 & 0.24 & 90 & 
$-3.201\pm0.011$ & $63.0\pm1.6$ \\ 
260-1550$\kms$ & Cep & 29 & 
$0.189\pm0.013$ & $0.212\pm0.014$ & $0.200\pm0.010$ & 
904 & 0.30 & 0.15 & 130 & 
$-3.198\pm0.012$ & $63.4\pm1.7$ \\ 
310-2000$\kms$ & SNe\,Ia & 20 & 
$0.175\pm0.021$ & $0.219\pm0.026$ & $0.192\pm0.016$ & 
1575 & 0.40 & 0.20 & 320 & 
$-3.220\pm0.019$ & $60.3\pm2.6$ \\ 
2000-10,000$\kms$ & TF clusters & 28 & 
$0.194\pm0.005$ & $0.197\pm0.005$ & $0.196\pm0.004$ & 
5089 & $<0.12$ & $<0.06$ & $<290$ & 
\nodata & \nodata \\ 
3000-20,000$\kms$ & SNe\,Ia & 62 & 
$0.192\pm0.003$ & $0.196\pm0.003$ & $0.194\pm0.002$ & 
7720 & $<0.15$ & $<0.07$ & $<550$ & 
$-3.213\pm0.004$ & $61.2\pm0.5$ \\
same with $\Lambda=0.7$ & & & & & & & & & &  
$-3.205\pm0.004$ & $62.3\pm0.5$ \\
% ***********************************************
\noalign{\smallskip}
\hline
\noalign{\smallskip}
% ***********************************************
\multicolumn{12}{l}{\footnotesize 
  based upon mean of $B$, $V$, and $I$ magnitudes}\\
% ***********************************************
\end{tabular}
\end{center}
% \end{table}
\end{sidewaystable}
% ***********************************************

     The main result from Table~\ref{tab:char} is the mean weighted
slope of the Hubble lines in Col.~6 from different distance
indicators. It amounts to $0.196\pm0.004$. This is impressively
close to the case of linear expansion with slope 0.2. It is stressed
again that the value of $H_{0}$ is therefore the same everywhere in the
free expansion field. $H_{0}$ can hence be determined at any distance
where the most suitable distance indicators are available. 
``Suitable'' means in this context high quality and a sufficient
quantity to reduce the random error caused by peculiar motions. The
influence of the latter is of course larger at small distances
requiring in that case a larger number of good distances. 

     The values $v_{\rm pec}$ in Col.~10 of Table~\ref{tab:char} hold
for field galaxies, but also include galaxies in groups because their
velocity dispersion is not significantly different. The few cluster
galaxies are entered with the mean cluster velocity. 
Even if the tabulated peculiar velocities carry statistical
errors of the order of $10-20\%$ there is no doubt that they increase
with distance. While the individual distances of 100 field and group
galaxies from the Hubble line give a mean value of 
$v_{\rm pec}=70\kms$ within $7\;$Mpc, $v_{\rm pec}$ increases 
to $130\kms$ at a distance of $900\kms$ (14.4$\;$Mpc). 
At still larger distances the contribution of the peculiar velocities
is of the same size as the distance errors and only upper limits can
be set for $v_{\rm pec}$. The upper limit of $v_{\rm pec}=290\kms$ at
a median velocity of $5000\kms$ seems realistic if it is compared with
the {\em three-dimensional\/} velocity of $460\kms$ (after subtraction
of the Virgocentric infall vector) of the entire Virgo complex
comprising a volume out to $\sim3000\kms$ with respect to the CMB
\citep{ST:85}.

% ******************************************************************
% 3. The zero-point calibration of TRGB, Cepheid, and SNeIa distances
% ******************************************************************
\section{The zero-point calibration of TRGB, Cepheid, and SNe\,Ia distances}
\label{sec:3}
In the previous section it was shown that the variation of the all-sky
value of $H_{0}$ with distance is unmeasurably small. For this
demonstration only relative distances were needed, yet for purely
practical purposes zero-pointed TRGB, Cepheid, and SNe\,Ia distances
were used. Their zero-point calibration follows now here.

% ********************************************************
% 3.1 The zero-point calibration of the TRGB 
% ********************************************************
\subsection{The zero-point calibration of the TRGB}
\label{sec:3:1}
\begin{figure}
   \centering
   \includegraphics[width=0.70\textwidth]{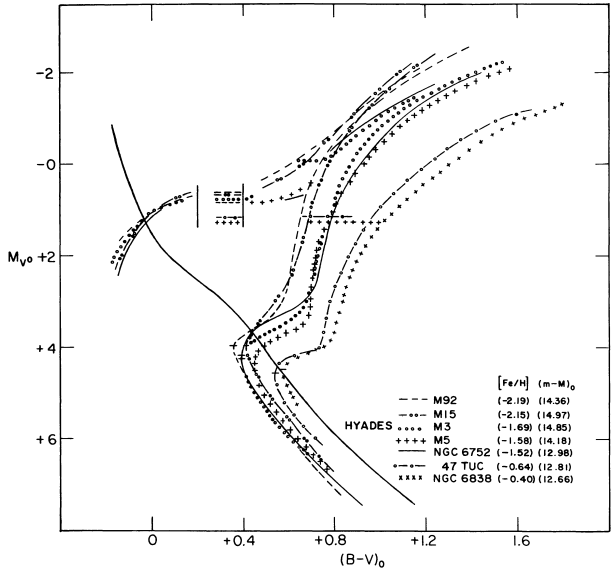}
    \caption{The composite CMD for seven globular clusters. 
      Note that the brightest red giant stars of the five most
      metal-poor clusters have very similar absolute magnitudes
      of about $M_{V}=-2.5$ (from \citealt{Sandage:86b}). The $I$
      magnitude of the brightest red giants is even more stable near
      $M_I=-4.05$ as found by \citet{DaCosta:Armandroff:90}.}
   \label{fig:CMD}
\end{figure}

When \citet{Baade:44a,Baade:44b}, using red-sensitive plates, pushed
to resolve the brightest stars in population~II galaxies such as
M\,32, NGC\,205, NGC\,147, and NGC\,185 he noticed that resolution
occurs abruptly upon reaching a fixed apparent magnitude. He explained
the sudden onset of resolution, later coined ``Baade's sheet'', as the
top of globular cluster like red-giant branches having approximately
constant luminosity. On modern plates the occurrence of Baade's sheet
is striking 
\citep[see e.g.][Panels 14, 15, 16, and 25]{Sandage:Bedke:94}.
The fixed luminosity of the brightest metal-poor giants was
theoretically explained by \citet{Rood:72} and
\citet{Sweigart:Gross:78} by their degenerate cores which make the
helium flash independent of mass, and it was observationally
confirmed when improved RR~Lyrae distances of globular clusters
allowed an alignment of their CMDs (Fig.~\ref{fig:CMD}). From early
beginnings as a distance indicator \citep{Sandage:71} Baade's sheet --
now named tip of the red-giant branch (TRGB) -- has become by now the
most powerful and most easily to use tool to determine distances out
to $\sim\!10\;$Mpc of galaxies containing an old population. The
development is marked by important papers by
\citet{DaCosta:Armandroff:90}, who introduced $I$ magnitudes for the
TRGB, \citet{Lee:etal:93}, \citet{Salaris:Cassisi:97}, and 
\citet{Sakai:etal:04}.

% ********************************************************
% Figure 2: CMD - TRGB
% ********************************************************
% ********************************************************

     The absolute $I$ magnitude of the TRGB was calibrated in
\citeauthor*{TSR:08} using 24 galaxies for which RR\,Lyrae distances
and apparent magnitudes $m^{\rm TRGB}_{I}$ are available. 
The latter were compiled from the literature and averaged where
necessary. The RR\,Lyrae distances are taken from Table~1 of 
\citeauthor*{TSR:08}, where also the original sources are referenced. 
The calibration for evolved RR\,Lyr stars is taken from
\citet[][Eq.~(8)]{ST:06}. The resulting TRGB luminosity is (omitting
Sag dSph and the Phoenix dwarf with less reliable observations)
\begin{equation}
M^{\rm TRGB}_{I}=-4.05\pm0.02
\label{eq:TRGB}
\end{equation}
for an old population with average metallicity [Fe/H]$_{\rm ZW}=-1.5$
in the system of \citet{Zinn:West:84}. The systematic error is
entirely determined by the RR\,Lyrae stars; it is estimated to be 
$\le0.1\mag$. It is stressed that the calibration is 
{\em independent\/} of any Cepheid distances.  

     The calibration in Eq.~(\ref{eq:TRGB}) agrees to better than
$0.1\mag$ with other results
\citep[e.g.][]{Bergbusch:VandenBerg:01,Sakai:etal:04,Bellazzini:etal:04,Rejkuba:etal:05}. 
\citet{Rizzi:etal:07} have fitted the Horizontal Branch (HB) of five
galaxies to the metal-dependent HB of \citet{Carretta:etal:00}
whose zero point rests on trigonometric parallaxes. Their result is
identical to Eq.~(\ref{eq:TRGB}) for the same average metallicity. 

     Model calculations show that the tip luminosity depends on
metallicity
\citep{Salaris:Cassisi:98,Bellazzini:etal:04,Rizzi:etal:07}. 
The sign of the change is not clear, however the authors agree that it
is not more than $\pm0.05\mag$ over the range of
$-2.0<\mbox{[Fe/H]}_{\rm ZW}<-1.2$; only for still higher
metallicities the tip magnitude is significantly fainter. 
The observational evidence fits into these results 
(see Fig.~1 of \citeauthor*{TSR:08}). The compromise here is to
adopt Eq.~(\ref{eq:TRGB}) throughout, independent of metallicity. 
The resulting error is certainly $<0.1\mag$ for red giants in the
quoted metallicity range. For many galaxies the tip metallicity (or
color) is not known;  the few cases which fall possibly outside this
wide metallicity range are statistically negligible.  

     For 240 galaxies with $I$ magnitudes of the TRGB in the
literature distance moduli (corrected for Galactic absorption) out to
$\sim\!10\;$Mpc are given in Table~\ref{tab:dist} Col.~6, on the uniform
basis of Eq.~(\ref{eq:TRGB}). The original sources are listed in Col.~11.

% ********************************************************
% 3.2 The P-L relation of Cepheids and their zero point
% ********************************************************
\subsection{The P-L relation of Cepheids and their zero point}
\label{sec:3:2}
Since \citet[][\citealt{Leavitt:Pickering:12}]{Leavitt:08} discovery
of the period-luminosity (P-L) relation of Cepheids it was assumed
that the P-L relation of classical Cepheids is universal. Hence
calibrated P-L relations in different wavelengths were derived 
\citep[e.g.][]{Kraft:61,ST:68,Madore:Freedman:91} and
indiscriminately applied. The assumption of universality, however, was
early on shattered when \citet{Gascoigne:Kron:65} found that the
Cepheids in LMC are bluer than those in the Galaxy -- which alone
precludes universal P-L relations -- and moreover when
\citet{Laney:Stobie:86} found the LMC Cepheids to be hotter than their
Galactic counterparts at given period. More recent data confirm the
dissimilarity of metal-rich Galactic Cepheids and metal-poor LMC
Cepheids.  

     Turning first to the Galactic Cepheids, good colors are available
for them mainly through the individual reddening corrections of
\citet[][\citealt{Fernie:etal:95}; 
slightly revised by \citealt{TSR:03}]{Fernie:90}. 
Distances are known of 33 Cepheids in clusters and associations
\citep{Feast:99}. Seven of the cluster distances have recently been
confirmed to within $0.1\mag$ by \citet{An:etal:07}. 
All cluster distances rest on an adopted Pleiades modulus of 5.61
which is secure to 0.02.  

     In addition absolute magnitudes of 36 Galactic Cepheids come from
the so-called BBW method \citep{Baade:26,Becker:40,Wesselink:46} of
moving atmospheres as improved by \citet{Barnes:Evans:76}. In 33 cases
the absolute magnitudes rest on radial-velocity measurements
\citep{Fouque:etal:03,Barnes:etal:03} and in three cases on
interferometric diameter measurements 
\citep[][and references therein]{Kervella:etal:04}. 
The 36 Cepheids and the cluster
Cepheids give quite similar slopes of their respective P-L relations
and agree at a period of $P = 10^{\rm d}$ to within $0.08\mag$. If the
two data sets are combined with equal weight they give the following
Galactic P-L relations in $B,V,I$ \citep{STR:04}:
\begin{eqnarray}
 \label{eq:gal:PL:B}
   M^{0}_{B} & = & -2.692\log P - 0.575 \\
 \label{eq:gal:PL:V}
   M^{0}_{V} & = & -3.087\log P - 0.914 \\
 \label{eq:gal:PL:I}
   M^{0}_{I} & = & -3.348\log P - 1.429.
\end{eqnarray}
They are adopted in the following. They give absolute magnitudes at
$P = 10^{\rm d}$ which are only $0.05\mag$ fainter than from
trigonometric {\em HST\/} parallaxes of 10 Cepheids
\citep{Benedict:etal:07} or $0.01\mag$ fainter if some Hipparcos
parallaxes are added \citep{vanLeeuwen:etal:07}. This excellent
agreement does not hold over the entire period interval as discussed
below.

     In a second step the LMC P-L relations can independently be
derived from 680 Cepheids with dereddened $B$, $V$, and $I$ magnitudes
from \citet{Udalski:etal:99}, to which 97 longer-period Cepheids are
added from various sources. They cannot be fitted by a single slope,
but show a break at $P = 10^{\rm d}$. The resulting LMC P-L relations
are \citep{STR:04} 
\begin{eqnarray}
\nonumber
  \mbox{for}~ \log P<1  &\qquad & \mbox{and for}~ \log P>1  \\ 
\label{eq:LMC:PL:B}
  M^0_{B} = -2.683\log P - 0.995&\qquad & M^0_{B} = -2.151\log P -1.404 \\
\label{eq:LMC:PL:V}
  M^0_{V} = -2.963\log P - 1.335&\qquad & M^0_{V} = -2.567\log P -1.634 \\
\label{eq:LMC:PL:I}
  M^0_{I} = -3.099\log P - 1.846&\qquad & M^0_{I} = -2.822\log P -2.084.
\end{eqnarray}
The zero point is set here by an adopted LMC modulus of 18.54. The
value is the mean of 29 determinations from different authors and
methods from 1997 to 2007 as compiled in \citeauthor*{STS:06} and
\citeauthor*{TSR:08}. Lower values in the literature come mostly from the
unjustified assumption that Galactic and LMC Cepheids are directly
comparable. -- The break at  $P=10^{\rm d}$ withstands several
statistical tests \citep{Ngeow:etal:05,Kanbur:etal:07,Koen:Siluyele:07}, 
besides being well visible by eye. Also the pulsation models of
\citet{Marconi:etal:05} show the break for the metallicity of LMC; it
is, however, absent for the higher metallicity of the Galaxy. 

     It is suggestive that the difference of the P-C and P-L relations
in the Galaxy and LMC is caused, at least in part, by the different
metallicity of the two galaxies. This leads to the following procedure
to derive Cepheid distances of galaxies with intermediate
metallicities. Two distances are derived for a given galaxy, one from
the Galactic and one from the LMC P-L relation. Noting that Galactic 
Cepheids have [O/H]$_{\rm Te}=8.62$ and LMC Cepheids 
[O/H]$_{\rm Te}=8.36$ -- in the [O/H]$_{\rm Te}$ scale of
\citet{Kennicutt:etal:03} and \citet{Sakai:etal:04} --  
the two distances are then interpolated and slightly extrapolated
according to the metallicity of the galaxy under study 
(\citeauthor*{STT:06}). 
The resulting Cepheid distances show no significant metallicity effect
if compared with TRGB, SNe\,Ia, and velocity distances
(\citeauthor*{TSR:08}). There are indications that eventually other
parameters like He-abundance \citep{Marconi:etal:05} must be involved
to explain all differences of the P-L relations.  

     The determination of Cepheid distances is complicated by the
necessity to deredden external Cepheids. This requires P-L relations
in at least two colors, which implies that an assumption on the
intrinsic color (P-C relation) must be made. Most Cepheids outside the
Local Group were observed with {\em HST\/} in $V$ and $I$
magnitudes. For distances derived from the LMC P-L relation in $V$ the
P-C relation must consistently be applied to derive $E(V\!-\!I)$. 
Distances derived from the Galactic P-L relation must correspondingly
be dereddened with the Galactic P-C relation. 
Since Galactic Cepheids are redder in $(V\!-\!I)$ than LMC Cepheids of
the same period, the reddening and the absorption corrections of a
Galactic Cepheid is therefore {\em smaller\/} than of an LMC Cepheid
of the same observed color and period. 
 
     The smaller absorption correction of the red, metal-rich Galactic
Cepheids is partially offset by the overluminosity of the blue,
metal-poor LMC Cepheids. 
As Eqs.~(\ref{eq:gal:PL:B})--(\ref{eq:LMC:PL:I}) show LMC Cepheids 
with $\log P=0.5$ are brighter in $B$, $V$, and $I$ than Galactic
Cepheids by $0.42$, $0.36$, and $0.30\mag$. The difference decreases
with increasing period and changes sign at about $\log P =1.5$  
(depending on wavelength).

% ***********************************************
%  Table 3: Distance difference Gal vs. LMC
% ***********************************************
\begin{table}
\begin{center}
\caption{Distance difference $\mu(\mbox{LMC}) - \mu(\mbox{Gal})$ of a
  Galactic Cepheid with period $P$ depending on whether it is reduced
  with the Galactic or LMC P-L and P-C relations.} 
\label{tab:03}
\footnotesize
\begin{tabular}{cccccccccc}
% ***********************************************
\hline
\hline
\noalign{\smallskip}
% ***********************************************
                                            & 
                                            & 
\multicolumn{2}{c}{Galaxy}                  & 
\multicolumn{2}{c}{LMC}                     & 
                                            & 
                                            & 
                                            & 
                                           \\
\multicolumn{1}{c}{$P$}                     & 
\multicolumn{1}{c}{$\log P$}                & 
\multicolumn{1}{c}{$M_{V}$}                 & 
\multicolumn{1}{c}{$(V\!-\!I)$}             & 
\multicolumn{1}{c}{$M_{V}$}                 & 
\multicolumn{1}{c}{$(V\!-\!I)$}             & 
\multicolumn{1}{c}{``$E(V\!-\!I)$''}        & 
\multicolumn{1}{c}{``$A_{V}$''}             & 
\multicolumn{1}{c}{``$M_{V}$''}             & 
\multicolumn{1}{c}{$\Delta(m\!-\!M)$}      \\ 
\multicolumn{1}{c}{(1)}                     & 
\multicolumn{1}{c}{(2)}                     & 
\multicolumn{1}{c}{(3)}                     & 
\multicolumn{1}{c}{(4)}                     & 
\multicolumn{1}{c}{(5)}                     & 
\multicolumn{1}{c}{(6)}                     & 
\multicolumn{1}{c}{(7)}                     & 
\multicolumn{1}{c}{(8)}                     & 
\multicolumn{1}{c}{(9)}                     & 
\multicolumn{1}{c}{(10)}                    \\
% ***********************************************
\noalign{\smallskip}
\hline
\noalign{\smallskip}
% ***********************************************
 5 & 0.70 & -3.07 & 0.676 & -3.49 & 0.613 & 0.063 & 0.21 & -3.28 & $+$0.21   \\
10 & 1.00 & -4.00 & 0.753 & -4.25 & 0.678 & 0.075 & 0.25 & -4.00 & $\pm$0.00 \\
15 & 1.18 & -4.56 & 0.799 & -4.66 & 0.752 & 0.047 & 0.15 & -4.51 & $-$0.05   \\
20 & 1.30 & -4.93 & 0.830 & -4.97 & 0.790 & 0.040 & 0.13 & -4.84 & $-$0.07   \\
25 & 1.40 & -5.24 & 0.355 & -5.23 & 0.821 & 0.034 & 0.11 & -5.12 & $-$0.12   \\
30 & 1.48 & -5.48 & 0.876 & -5.42 & 0.846 & 0.030 & 0.10 & -5.38 & $-$0.15   \\
% ***********************************************
\noalign{\smallskip}
\hline
\noalign{\smallskip}
% ***********************************************
% ***********************************************
\end{tabular}
\end{center}
\end{table}
% ***********************************************

     Table~\ref{tab:03} shows the effect on distance if an unreddened
Galactic Cepheid with period $P$ and Galactic properties is 
``mistreated'' with the $V$ and $I$ P-L relations of LMC. 
Cols. 3 and 4 give $M_{V}$ and $(V\!-\!I)$ for a Galactic Cepheid, 
Cols. 5 and 6 the same for an LMC Cepheid. If the latter values are
applied to a Galactic Cepheid one derives the spurious reddenings and
absorptions in Cols. 7 and 8. The absorption diminishes the effective
LMC luminosity in Col.~5 to the values in Col. 9. A comparison of
Col.~9 with Col.~3 gives then the distance error in the sense
$\mu(\mbox{LMC})- \mu(\mbox{Gal})$. The change of sign of the distance
error with period makes that a Cepheid sample with a wide period
distribution will be assigned a rather reasonable {\em mean\/}
distance. But most Cepheids outside the Local Group have long periods
($P_{\rm median} \approx 25^{\rm d}$) and, if metal-rich, their
distances will be systematically underestimated by $\sim\!0.1\mag$, or
even more in case of very metal-rich Cepheids with particularly long
periods.   

     The steep P-L relations of the Galaxy are shared by the
metal-rich Cepheids of some other galaxies (NGC\,3351, NGC\,4321, M\,31), 
and there is a general trend for less metal-rich Cepheids to exhibit
progressively flatter slopes (\citeauthor*{TSR:08}, Fig.4). This
supports the interpretation that metallicity is at least one of the
parameters that determines the P-L slope. 
But the metal-rich Cepheids in an inner field of NGC\,4258
\citep{Macri:etal:06} define a P-L slope as flat as in LMC. It follows
from this that still another parameter than metallicity affects the
P-L relations. The models of \citet{Marconi:etal:05} identify the He
content as a prime candidate.

     The difference of the P-L relations in the Galaxy and in LMC
cannot be questioned, but the Galactic slope, resting on only 69
open-cluster and BBW calibrators, may still be open to revisions.
\citet{Gieren:etal:05} and \citet{Fouque:etal:07} have in fact
proposed less steep slopes by changing in case of the BBW method the
period dependence of the projection factor $p$, which converts
observed radial velocities into pulsational velocities. 
Also \citet{Benedict:etal:07} and \citet{vanLeeuwen:etal:07} plead for
a flatter slope on the basis of a dozen parallax measurements. 
However, one must then discard the evidence of cluster Cepheids. In
any case the assumption of {\em one\/} universal flat, LMC-like P-L
relation would leave unexplained the redness of the Galactic Cepheids
and the break of the LMC P-L relation at $P=10^{\rm d}$ and its
absence in the Galaxy. 

     The absorption-corrected distance moduli of 37 galaxies, adjusted
for metallicity as described above, were derived by 
\citeauthor*{STT:06} and of four additional galaxies by
\citeauthor*{TSR:08}, where also the original sources are given. 
The Cepheids of three very metal-poor galaxies were tied without
further metallicity corrections to those of SMC for which a mean
modulus of $\mu_{\rm SMC}=18.93\pm0.02$ was adopted from five 
independent methods (see \citeauthor*{TSR:08}, Table~7). 
The total of 43 Cepheid distances is compiled in Table~\ref{tab:dist},
Col.~7. The 29 galaxies with distances $> 4.4\;$Mpc are shown in a
distance-calibrated Hubble diagram (Fig.~\ref{fig:hubble}b). 
The slope of the Hubble line has been discussed in 
Sect.~\ref{sec:3:2} without the necessity of zero-pointed distances. 
With the calibration now in hand the intercept becomes 
$a=-3.198\pm0.012$ (Table~\ref{tab:char}).
 
     The random error of the Cepheid distances will be discussed in
Sect.~\ref{sec:3:4}. For a $10^{\rm d}$ Cepheid with Galactic
metallicity the {\em systematic\/} error of the distance, which
depends on cluster Cepheids, BBW distances, and which agrees so well
with trigonometric parallaxes, is not more than $0.05\mag$. For other
metallicities the distance error may increase with 
$\Delta\mu=(0.05\pm0.10)\Delta\mbox{[O/H]}_{\rm Te}$ as shown from a
comparison of Cepheid distances with TRGB, SNe\,Ia, and velocity
distances (\citeauthor*{TSR:08}). The dependence is insignificant and
will in any case, even for the lowest metallicities, introduce an
additional distance error of less than $0.1\mag$.

% ********************************************************
% 3.3 The zero-point calibration of SNeIa
% ********************************************************
\subsection{The zero-point calibration of SNe\,Ia}
\label{sec:3:3}
The luminosity calibration of SNeIa was discussed in detail by 
\citeauthor*{STT:06} and is not repeated here. For 10 normal SNe\,Ia,
corrected for Galactic and internal absorption and homogenized to a
common decline rate and color, Cepheid distances are available. They
yield the following absolute magnitudes at $B$ maximum
(\citeauthor*{STS:06}):  
\begin{equation}
M_{B}= -19.49\pm0.04 \qquad M_{V}=-19.46\pm0.04 \qquad M_{I}=-19.22\pm0.04. 
\label{eq:SNeIa}
\end{equation}
They are brighter by $0.12\mag$ than adopted by
\citet{Freedman:etal:01} and by $0.25\mag$ than derived from only four
calibrators by \citet{Riess:etal:05}. A strict comparison of these
values is not possible because the magnitudes are reduced to standard
decline rates and colors, but the fainter values are based on a
version of the P-L relation adopted for the metal-poor LMC Cepheids,
although most of the calibrators are metal-rich. Since most of the
relevant Cepheids have also long periods the difference in metallicity
is important (cf. Table~\ref{tab:03}).

     A first attempt to independently calibrate SNe\,Ia through the
TRGB rests so far on only two galaxies with their own TRGB distances
and on two more galaxies in the Leo~I group, for which a mean TRGB
distance can be used. The quite preliminary result is
$M_{V}=-19.37\pm0.06$ (\citeauthor*{TSR:08}) 
which is in statistical agreement with Eq.~(\ref{eq:SNeIa}). As more
TRGB distances to SNe\,Ia will become available the method will become
highly competitive. 

     If Eq.~(\ref{eq:SNeIa}) is combined with the consistently reduced
apparent magnitudes in $B$, $V$, and $I$ of 98 normal SNe\,Ia from
\citet{RTS:05} one obtains their true distance moduli. The
sample has been divided into two subsets. The one comprises the 22
SNe\,Ia with $v_{220}<2000\kms$ already discussed in 
Sect.~\ref{sec:2:4}.
They define the distance-calibrated Hubble diagram in
Fig.~\ref{fig:hubble}c and an intercept
of $-3.220\pm0.019$ which is shown in Table~\ref{tab:char}. The
more distant subset contains the 62 SNeIa with 
$3000 < v_{\rm CMB} < 20,000\kms$. 
They yield an intercept of $a=-3.205\pm0.004$ after allowance for
$\Omega_{\Lambda}=0.7$. (For a flat Universe with $\Omega_{M}=0.3$,
$\Omega_{\Lambda}=0.0$, the intercept becomes $a=-3.213\pm0.004$,
cf. Table~\ref{tab:char}).   

     The intercept of the Hubble line cannot be compared with the one
obtained by \citet{Jha:etal:07}, because the apparent SN\,Ia
magnitudes were normalized in a different way and reduced to different
standard parameters than in \citet{RTS:05}. The same holds for the
work of \citet{Wang:etal:06}. They obtain from 73 SNe\,Ia a Hubble
diagram with a dispersion of only $\sigma_{m}=0.12$ 
in $V$ and derive a value of $H_{0}=72.1\pm1.6$ (statistical error)
using low Cepheid distances for their calibrating SNe\,Ia.
However, if the Cepheid distances in Table~\ref{tab:dist} are used for
their calibrators one finds $H_{0}=65.4\pm1.5$. The 5\% difference
from our preferred value reflects the uncertainties caused by the
dereddening and normalization of the observed SN\,Ia magnitudes.   

     The intercepts $a$ obtained from the zero-point calibration of the
TRGB, Cepheid, and SN\,Ia distances are collected in
Table~\ref{tab:char}, Col.~11.

% ********************************************************
% 3.4 Comparison of different distance determinations
% ********************************************************
\subsection{Comparison of different distance determinations}
\label{sec:3:4}
%
% **********************************************
% 3.4.1. Comparison of individual galaxies
% **********************************************
\subsubsection{Comparison of individual galaxies}
\label{sec:3:4:1}
The internal accuracy of the TRGB and Cepheid distances in
Table~\ref{tab:dist} can be determined by comparison with RR\,Lyrae
distances and by intercomparison (Table~\ref{tab:04}).  
 
% ***********************************************
%  Table 4: Comparison of different distance determinations.
% ***********************************************
\begin{table}
\begin{center}
\caption{Comparison of different distance determinations.} 
\label{tab:04}
\footnotesize
\begin{tabular}{lcrc}
% ***********************************************
\hline
\hline
\noalign{\smallskip}
% ***********************************************
                                      &
\multicolumn{1}{c}{N}                 & 
\multicolumn{1}{c}{$\Delta\mu$}       & 
\multicolumn{1}{c}{$\sigma_{m-M}$}    \\
% ***********************************************
\noalign{\smallskip}
\hline
\noalign{\smallskip}
% ***********************************************
$\mu_{\rm TRGB}-\mu_{\rm RRLyr}$ & 20 & $ 0.00\pm0.02$ & 0.08 \\
$\mu_{\rm Cep}-\mu_{\rm TRGB}$   & 17 & $-0.05\pm0.03$ & 0.13 \\
% ***********************************************
\noalign{\smallskip}
\hline
\noalign{\smallskip}
% ***********************************************
\end{tabular}
\end{center}
\end{table}
% ***********************************************

     The zero difference of the TRGB and RR\,Lyr distances is no
surprise because the latter have served as calibrators. More
remarkable is the small dispersion which implies that the random error
of either distance indicator is certainly less than $0.1\mag$. A
generous error of $0.15\mag$ has been adopted above. Still more
remarkable is in view of the {\em independent\/} zero points the
barely significant difference of $0.05\pm0.03\mag$ between the Cepheid
and TRGB distances, the former being smaller. The difference is
neglected because it is not seen in the intercepts $a$ 
(Table~\ref{tab:char}), which involve a larger number of galaxies. 
The dispersion of $0.13\mag$ between the two distance indicators sets
again an upper limit of say $0.15\mag$ for the random error of the
Cepheid distances. Also the SN\,Ia distances carry a random error of
not more than $0.15\mag$ as seen from the dispersion of the Hubble
diagram of the distant SNe\,Ia.
 
     There is only a limited number of galaxies with independent
distances of comparable accuracy and with presumably small systematic
errors. One case is NGC\,4258 for which \citet{Herrnstein:etal:99} have 
determined a modulus of $29.29\pm0.10$ from the Keplerian motion of
water maser sources about the galaxy center; the value is in
statistical agreement with $29.41\pm0.11$ from the mean of the TRGB
and Cepheid distance. \citet{Ribas:etal:05} have derived the distance
of NGC\,224 (M\,31) from an eclipsing binary to be $24.44\pm0.12$ in
perfect agreement with the mean RR\,Lyr, TRGB, and Cepheid distance. 
The eclipsing binary distance of NGC\,598 (M\,33) of $24.92\pm0.12$ by 
\citet{Bonanos:etal:06} is only marginally larger 
than the mean of $24.69\pm0.09$ from the RR\,Lyr stars, the TRGB, and the
Cepheids. Interesting are also the four Cepheid distances that
involve near-infrared magnitudes in $J$ and $K$, which are believed to
be less susceptible to metallicity effects and which are tied to the
$J$,$K$ P-L relation of LMC by \citet{Persson:etal:04}. 
The distances of 
NGC\,300 \citep{Rizzi:etal:06}, 
NGC\,3109 \citep{Soszynski:etal:06}, 
NGC\,6822 \citep{Gieren:etal:06}, and 
IC\,1613 \citep{Pietrzynski:etal:06}
differ on average from the independent distances in Table~\ref{tab:dist}
by only $0.00\pm0.04$ if $(m-M)^{0}_{\rm LMC}=18.54$ is adopted.  

     From this it seems that the distances in Table~\ref{tab:dist} form
a {\em homogeneous\/} system based on a common zero point. The random 
distance error is probably $\le0.15\mag$ for a galaxy with one
distance determination and accordingly smaller in cases of two and
three determinations. Table~\ref{tab:dist} is therefore believed to be
the best net of local distances presently available. 
It comprises a wide range of galaxy types; normal E/S0
galaxies with $v_{220}<1000\kms$, however, are painfully missing.

% **********************************************
% 3.4.2. Comparison of the intercept a
% **********************************************
\subsubsection{Comparison of the intercept $a$}
\label{sec:3:4:2}
The most interesting result of the previous Section is the close
agreement of the intercepts $a$, as compiled in Table~\ref{tab:char},
Col.~11, from the Population~II (old stars) TRGB distances larger than 
4.5$\;$Mpc and from the young-Population~I Cepheid distances, because
they rest on independent zero points. The difference of 
$\Delta a = 0.003\pm0.016$ (corresponding to $0.02\pm0.08\mag$) is as
good as could be expected and reflects on the quality of the mutual
zero-point calibrations. One could object that the agreement is
coincidental because the median distance of the Cepheids is 2.4 times
larger than that of the TRGB galaxies, but the invariance of $H_{0}$
with distance is just what was predicted in Sect.~\ref{sec:2} from
only the slopes of the different Hubble diagrams. 

     To include also the weight of the numerous nearby and distant
SNe\,Ia (in the latter case with allowance for 
$\Omega_{\Lambda} = 0.7$) their $a$-values were averaged with the one
from Cepheids to give $a =-3.210\pm0.012$. The SNe\,Ia cannot improve
the zero point since they are calibrated with a subset of the same
Cepheids, but they help to decrease the statistical error and directly
lead into the large-scale expansion field. The preferred solution here
is the mean of the latter Cepheid-based value of $a$ and
$a=-3.201\pm0.011$ from the independent TRGB galaxies, 
i.e.\ $a=-3.205\pm0.09$. 

From Eq.~(\ref{eq:04}) follows then that 
\begin{equation}
   H_{0}(\mbox{on all scales})= 62.3\pm1.3 (\mbox{statistical
     error})\; \pm4 (\mbox{systematic error}). 
\label{eq:H0}
\end{equation}
The systematic error here is estimated in the following way. A 10\%
error could be explained only if 1) $H_{0}$ varied noticeably with
distance which is excluded by the slope of the Hubble line very close
to $0.2$ (Sect.~\ref{sec:2}), or 2) if the adopted zero points of the
TRGB and of Cepheids were both changed in the same direction by
$0.2\mag$, which seems impossible. Therefore the systematic error is
still rather pessimistically estimated to be 6\%.   
It may be noted that omission of the $220\kms$ Virgocentric infall
correction would decrease the {\em local\/} value of $H_{0}$ by 
$\sim\!5$ units.

% ******************************************************************
% 4. Additional distance indicators
% ******************************************************************
\section{Additional distance indicators}
\label{sec:4}
Too many proposals have been made, how to measure galaxy distances, to
do justice to them here. Only a few methods are mentioned which have
been used widely and which have provided sufficient distances for
statistical tests.

% ********************************************************
% 4.1 21cm line widths
% ********************************************************
\subsection{21cm line widths Tully-Fisher (TF) method}
\label{sec:4:1}
\begin{figure}
   \centering
   \includegraphics[width=0.95\textwidth]{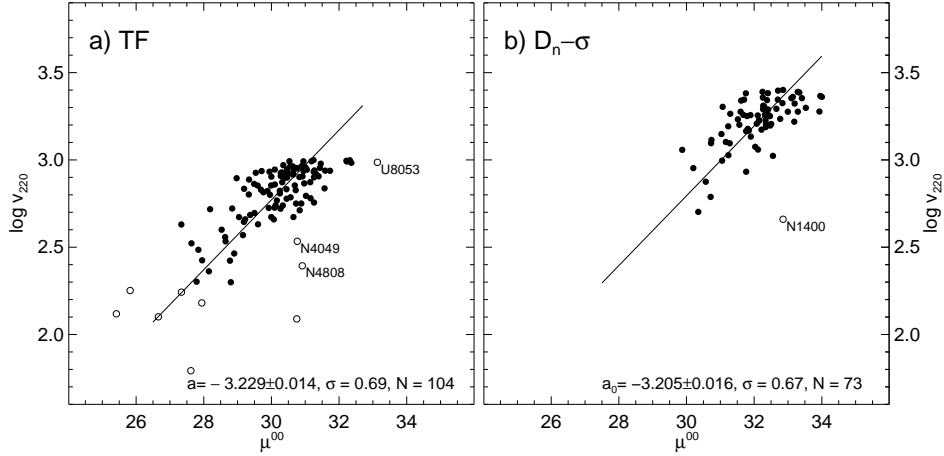}
    \caption{The Hubble diagram of a) TF distances of a complete
      sample of spiral galaxies with $v_{220}<1000\kms$, b)
      $D_{n}\!-\!\sigma$ distances of E galaxies with $v_{220}<2500\kms$,
      $H_{0}=62$ assumed. The open symbols are galaxies with
      $\mu^{00}<28.2$ and some outliers. The apparent widening of the
      Hubble line with distance is a statistical effect due to
      relatively large distance errors.}
    \label{fig:tf}
\end{figure}

The spectral line width of the 21cm line or of optical lines 
\citep[see][]{Mathewson:etal:92a}, corrected for inclination $i$, are
a measure of the rotation velocity of spirals and hence correlate with
galaxy mass and luminosity \citep{Gouguenheim:69}. The relation has
been applied by \citet{Tully:Fisher:77} and many subsequent authors
(some of which are quoted in \citealt{Tammann:06}) for the distance
determination of spirals. A reliable rotation velocity requires 
$i<45^{\circ}$ which unfortunately implies large corrections for
internal absorption. A Hubble diagram of a  {\em complete\/}
distance-limited sample of 104 inclined spirals with  
$v_{220}<1000\kms$ from \citet{Federspiel:99} gives the Hubble diagram
shown in Fig.~\ref{fig:tf}a. The scatter of $\sigma_{m}=0.69$ is very
large, too large in fact to define an independent slope of the Hubble
line. Even the assumption that peculiar velocities contribute 
$\sigma_{m}=0.30-0.40$ leaves an intrinsic scatter of
$\sigma_{m}=0.55$. This invites in case of flux-limited samples large
selection effects and too large values of $H_{0}$ as well as too small
estimates of the intrinsic dispersion. With the zero point from 31
Cepheids (\citeauthor*{STS:06}) one obtains for the distance-limited
sample $H_{0}=59.0\pm1.9$. This result, depending directly on the
Cepheid calibrations, is statistically different from the result of the
Cepheids themselves, which reveals some of the intricacies of the
method.  
 
     With the above calibration one obtains from a {\em complete\/}
sample of 49 inclined, untruncated Virgo cluster spirals, as compiled
by \citet{Federspiel:etal:98}, and after a small correction for the
color difference between calibrators and cluster galaxies a mean TF
distance of $\mu^{0}=31.58\pm0.16$, or reduced to the center of the
Local Group $\mu^{00}=31.62$ (\citeauthor*{STS:06}). -- 
\citet{Tully:Pierce:00} have derived for an almost complete sample of
38 inclined spirals of the UMa cluster with $B$, $R$, $I$, and $K'$
photometry $\mu^{0}=31.35\pm0.06$. 
After recalibrating their 24 calibrators with the present Cepheid
distances one obtains $\mu^{0}=\mu^{00}=31.45$. However, the UMa field
is complex and may be divided into two groups at slightly different
distances giving moduli of $31.26\pm0.16$ for UMa~I and 
$31.58\pm0.17$ for UMa~II \citep{Sandage:08} --
The Fornax cluster with only few inclined spirals does not yield well
to the TF method.   

% ********************************************************
% Figure 3: 2 additional Hubble diagrams [TF + Dn-sig]
% ********************************************************
% ********************************************************

% ********************************************************
% 4.2 Dn-sigma
% ********************************************************
\subsection{$D_{n}\!-\!\sigma$ or the fundamental plane (FP)}
\label{sec:4:2}
The correlation of the velocity dispersion $\sigma$ of E galaxies with
their luminosity was pointed out by \citet{Minkowski:62} and
\citet{Faber:Jackson:76}. Later the luminosity was replaced by a
suitably normalized diameter $D_{n}$ \citep{Dressler:etal:87} or by
surface brightness \citep{Djorgovski:Davis:87}. The method was
extended to bulges of spiral galaxies by \citep{Dressler:87} who
derived $H_{0}=67\pm10$. \citet{Faber:etal:89} have presented a wealth
of $D_{n}-\sigma$ measurements from which they have derived relative
distances $R_{e}$. A subset of 73 of their galaxies brighter than
$13.5\mag$ and with $v_{220} < 2500\kms$ constitute not a strictly
complete, but apparently a quite fair sample. Their Hubble diagram is
shown in Fig.~\ref{fig:tf}b. The data do not allow to determine the
slope, but a forced slope of 0.2 is acceptable. The large observed
scatter of $\sigma_{m}$ is about the same as for the TF method. 
Since no primary calibrators are available for E galaxies a value of 
$H_{0}=62$ is {\em assumed}. This leads to the following calibration    
\begin{equation}
   \mu^{0} = 5\log R_{\rm e} + 15.93.
\label{eq:FP}
\end{equation}
If this relation is applied to the 15 Virgo cluster members of the
sample one obtains $\mu^{0}=31.56\pm0.10$, which is still useful
because it is independent of the cluster velocity. The corresponding
mean of the 10 E galaxies in the sample, which are members of the
Fornax cluster, give $\mu^{0}=31.69\pm0.16$.

% ********************************************************
% 4.3 Other distance indicators
% ********************************************************
\subsection{Other distance indicators}
\label{sec:4:3}
\paragraph{Surface Brightness Fluctuations (SBF).} Surface brightness
fluctuations of E/S0 galaxies as a measure of distance have been
introduced by \citet{Tonry:Schneider:88} and have been applied with
variable success (references in \citealt{Tammann:06}). One of the
difficulties of the method is, as in case of the $D_{n}\!-\!\sigma$
method, that no primary calibration for E galaxies exists, and S0
galaxies may or may not follow the same relation and may be more
susceptible to dust. The 123 SBF distances compiled by 
\citet{Tonry:etal:01} give a Hubble diagram with somewhat 
less scatter ($\sigma_{m}=0.55$) than from TF or $D_{n}\!-\!\sigma$
distances, but the slope is {\em significantly\/} steeper than
0.2. This proves the SBF scale to be compressed with $H_{0}$ increasing
spuriously with distance. The problem could be caused by selection
effects, but rather it is inherent to the method. 
The careful work of \citet{Mei:etal:07} on Virgo cluster ellipticals
does not (yet) contribute to the determination of $H_{0}$ because they
{\em assume\/} a mean cluster distance.   

\paragraph{Planetary nebulae (PNLF).} Following a proposal of 
\citet{Ford:Jenner:78}
the luminosity function of the shells of planetary nebulae in the
light of the [OIII]$\lambda5007$ line has been used as a distance
indicator. But the maximum luminosity seems to depend on population
size \citep{Bottinelli:etal:91,Tammann:93}, chemical composition
and age \citep{Mendez:etal:93,Ciardullo:etal:02}, 
and dynamics \citep{Sambhus:etal:05}. About 30 galaxies,
mainly from \citet{Ciardullo:etal:02}, with PNLF distances $>28.2$
define a Hubble diagram with large scatter and steep slope implying 
$H_{0}$ to increase outwards. At $\sim\!1000\kms$ the PNLF distance
scale has lost about 0.5 mag as shown by five galaxies
\citep{Feldmeier:etal:07}  
with known SNe\,Ia whose resulting mean luminosity of 
$M_{V}(\mbox{SNe\,Ia})=-18.96$ should be compared with 
Eq.~(\ref{eq:SNeIa}).  

\paragraph{Luminosity classes (LC).} The luminosity of a spiral galaxy
correlates with the ``beauty'' of its spiral structure. Correspondingly 
spirals were divided into class I (the brightest) to V (the faintest)
by \citet{vandenBergh:60a,vandenBergh:60b,vandenBergh:60c}
with additional galaxies classified and modified by
\citet{Sandage:Bedke:94}. The purely morphological classification
is independent of distance; it yields therefore relative distances
which were valuable for many years when velocity distances were
suspected to be severely distorted by peculiar and streaming motions,
but the dispersion is large which makes the method susceptible to
bias. Locally calibrated and bias-corrected distances led to values of
$H_{0}$ near 55 \citep{ST:75b,Sandage:99}.
\newline

     Some methods like the brightest blue stars, used extensively by
Hubble, and the size of the largest HII regions \citep{Sersic:59} have
lost their former importance as distance indicators. Others show
increasing potential like novae which may reach out to the Virgo
cluster \citep{Gilmozzi:DellaValle:03}, but it is difficult to
determine an independent zero point for them and they require much
telescope time.     
-- The turnover magnitude of the luminosity function of globular
clusters (GCLF) was proposed as a standard candle by
\citet{vandenBergh:etal:85}. The luminosity of the turnover was
calibrated using RR\,Lyr distances in the Galaxy and the Cepheid
distance of M\,31, to be $M^{\rm TO}_{V}=-7.62$ 
\citep[][see also \citealt{DiCriscienzo:etal:06}]{ST:95}.
A simple-minded application to two galaxies in the Leo group and eight
galaxies in the Virgo cluster gave distances that agree with those
adopted here (Table~\ref{tab:groups} \& \ref{tab:virgo}) to within
$\sim\!0.1\mag$ \citep{TS:99}.
\citet{Kavelaars:etal:00} found from the same method the Coma cluster
to be more distant than the Virgo cluster by $4.06\pm0.11$; this leads
with $(m-M)_{\rm Virgo}=31.60\pm0.08$ (from Table~\ref{tab:virgo}) to 
$(m-M)_{\rm Coma}=35.66\pm0.14$. However, the simple application of
the GCLF method is questioned by the bimodal and varying color and
luminosity distribution of the GCs in different galaxies
\citep{Larsen:etal:01}. 

     Some ``physical'' distances do not make use of any known
astronomical distance, but are derived from the physics or geometry of
an object. Some are mentioned elsewhere in this paper, like 
BBW distances \citep{Fouque:etal:03}, 
eclipsing binaries (\citealt{Ribas:etal:05}; see also \citealt{Ribas:07}), 
the water maser distance of NGC\,4258 \citep{Herrnstein:etal:99}, and
the luminosities of Cepheids \citep{Marconi:etal:05}. 
The light echo distance of SN\,1987A \citep{Panagia:05} has been
incorporated into the zero-point distance of LMC. 
Much work has been devoted to model the luminosities of SNe\,Ia 
\citep[for a summary see][]{Branch:98}. The SN\,II models of
\citet{Eastman:etal:96} give distances which lead to an unrealistic
increase of $H_{0}$ with distance. Models of typeII-P SNe by 
\citet{Nugent:etal:06} give a mean value of $H_{0}=67\pm4$ for 19
objects, while \citet{Hamuy:Pinto:02} find $H_{0}=55\pm12$ for eight
objects. \citet{Nadyozhin:03} has derived from a refined model for the
same objects $H_{0}=55\pm5$, but the result is still quite sensitive to
the input parameters \citep{Blinnikov:etal:05}. 
The list of physical distance determinations could be much
extended, but it is a typical problem that their systematic errors are
difficult to determine and that they are often restricted to one or a
few objects.  
 
     Physical methods to determine $H_{0}$ at large distances have the
disadvantage to depend on the cosmological model. Important results
will eventually come from the Sunyaev-Zeldovich effect (SZE) of X-ray
clusters, but with values of $H_{0}=59-77$ and systematic errors of
$\sim\!20\%$ the results are not yet useful
\citep{Udomprasert:etal:04,Jones:etal:05,Bonamente:etal:06}.
-- A powerful method to measure large distances comes from
gravitational lensed quasars, however the
solution for $H_{0}$ is sensitive to the mass distribution of the
lens, to dark halos and companion galaxies, and even to the
large-scale structure in front of the lens and behind. Recent results
are $H_{0}~\sim70$ \citep{Fassnacht:etal:06} and
$H_{0}=64^{+8}_{-5}$ \citep{Read:etal:07} if $\Omega_{M}=0.3$,
$\Omega_{\Lambda}=0.7$ is assumed. 
\citet{Auger:etal:07} can fit the source SBS 1520+530 with $H_{0}=72$
if a steep mass profile of the lens is adopted, but an isothermal
model gives $H_{0}\approx46$.

     The acoustic fluctuation spectrum of the WMAP3 data is interpreted
to give a value of $H_{0}=72$ \citep{Spergel:etal:07}, which is also
consistent with the red giant galaxy distribution of the Sloan Digital
Sky Survey \citep{Tegmark:etal:06}. However, the result is
model-dependent, a priori assuming for instance a perfectly flat
Universe or a static value of the parameter $\Lambda$. 
A fundamentally different model allows for time dilation effects and
gives a proper integration over voids and filaments by introducing
density fluctuations into the Einstein equations as they affect
$H_{0}$, $\Lambda$, and the putative, but here illusory acceleration
\citep{Wiltshire:07a,Wiltshire:07b}.
This model gives a best-fit value of $H_{0}=61.7\pm1.2$
\citep{Leith:etal:08}.

% ******************************************************************
% 5. Distances of groups and clusters
% ******************************************************************
\section{Distances of groups and clusters}
\label{sec:5}
\begin{figure}
   \centering
   \includegraphics[width=0.55\textwidth]{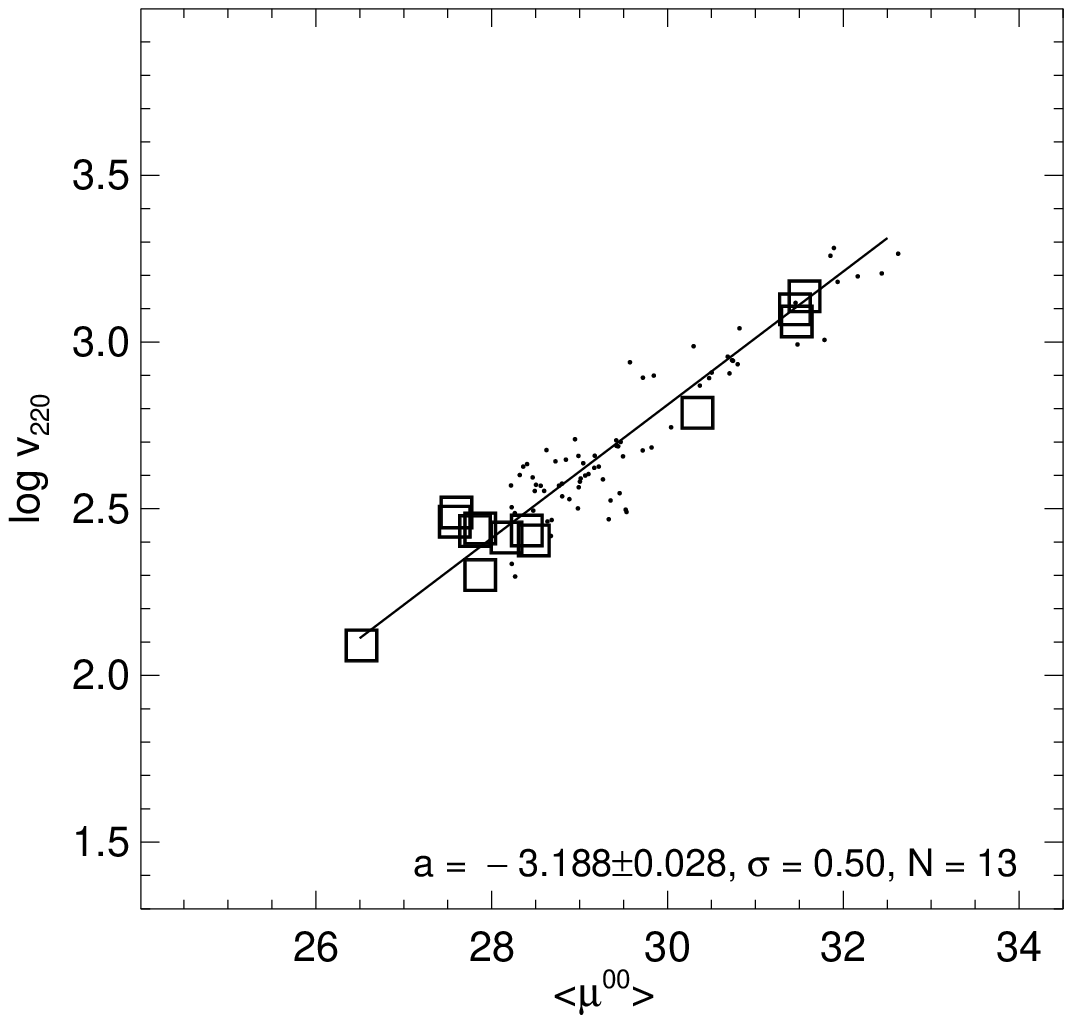}
    \caption{The Hubble diagram of 10 groups and of the UMa, Virgo,
   and Fornax clusters. Field galaxies which are not assigned to a
   group are shown with dots.}
    \label{fig:gc}
\end{figure}

The galaxies in Table~\ref{tab:dist} are assigned to different groups in
Col.~2. If the distances $\mu^{00}$ and velocities within a given group
are averaged with equal weight one obtains the values shown in
Table~\ref{tab:groups}. In addition the data for the distances of the UMa,
Virgo, and Fornax clusters are compiled in Table~\ref{tab:virgo} where
also the evidence from the TF and $D_{n}\!-\!\sigma$ method is
included. The Hubble diagram of the groups and clusters is shown in
Fig.~\ref{fig:gc}.  
A free fit of the Hubble line, including objects as close as
$3.3\;$Mpc~(!), gives a slope of $0.181\pm0.017$. 
A forced fit with slope 0.2 gives $H_{0}=64.8\pm4.2$ or, excluding the
deviating cases of the IC\,342 and NGC\,2403 groups,
$H_{0}=60.4\pm2.5$.
The average deviation from the Hubble line is only $55\kms$ without a
clear trend to depend on distance. Local groups and clusters follow
hence, after allowance for a Virgocentric flow model, a quiet
Hubble flow.

     The 72 galaxies of Table~\ref{tab:dist} with $\mu^{00}>28.2$, which
are {\em not\/} assigned to a group or cluster, have about the same
dispersion about the Hubble line as the groups and clusters. They give
$H_{0}=63.1\pm1.6$. 

     The distance of the Coma cluster can be estimated from its
relative distance to the Virgo cluster. The difference
$\Delta(m-M)_{\rm Coma - Virgo}$ is 3.74 from the $D_{n}-\sigma$
method \citep{Faber:etal:89} and 4.06 from globular clusters
\citep{Kavelaars:etal:00}. Adding the mean to the Virgo modulus in
Table~\ref{tab:virgo} gives $(m-M)_{\rm Coma}=35.50\pm0.15$. The
cosmic recession velocity of the Coma cluster, freed of all non Hubble
velocities, can be inferred from $D_{n}-\sigma$ distances relative to
Coma of nine distant clusters \citep{Jorgensen:etal:96} to be
$7800\pm\kms$ \citep[][Fig.~7]{Tammann:06}, from which follows
$H_{0}=62.0\pm5.0$.  

% ***********************************************
%  Table 5: Distances of groups
% ***********************************************
\begin{table}
\begin{center}
\caption{Distances of groups.} 
\label{tab:groups}
\footnotesize
\begin{tabular}{lrrcc}
% ***********************************************
\hline
\hline
\noalign{\smallskip}
% ***********************************************
\multicolumn{1}{c}{Group}                    &
\multicolumn{1}{c}{N}                        & 
\multicolumn{1}{c}{$\langle v_{220}\rangle$} & 
\multicolumn{1}{c}{$\langle\mu^{0}\rangle$}  &
\multicolumn{1}{c}{$\langle\mu^{00}\rangle$} \\
\multicolumn{1}{c}{(1)}                      & 
\multicolumn{1}{c}{(2)}                      & 
\multicolumn{1}{c}{(3)}                      & 
\multicolumn{1}{c}{(4)}                      & 
\multicolumn{1}{c}{(5)}                      \\ 
% ***********************************************
\noalign{\smallskip}
\hline
\noalign{\smallskip}
% ***********************************************
M\,31     & 15 & -21 & 24.39 & 21.73 \\
Scl1      & 4  & 123 & 26.52 & 26.50 \\
M\,81     & 23 & 200 & 27.87 & 27.76 \\
NGC\,4236 & 2  & 254 & 28.48 & 28.46 \\
CVn       & 21 & 259 & 28.17 & 28.21 \\
Scl2      & 6  & 271 & 27.81 & 27.68 \\
M\,83     & 10 & 272 & 28.40 & 28.63 \\
CenA      & 28 & 276 & 27.87 & 28.16 \\
IC\,342   & 7  & 289 & 27.58 & 27.30 \\
NGC\,2403 & 7  & 308 & 27.60 & 27.43 \\
Leo\,I    & 7  & 613 & 30.34 & 30.39 \\
% ***********************************************
\noalign{\smallskip}
\hline
\noalign{\smallskip}
% ***********************************************
\end{tabular}
\end{center}
\end{table}
% ***********************************************

% ***********************************************
%  Table 6: Cluster distances mu^00
% ***********************************************
\begin{table}
\begin{center}
\caption{Cluster distances $\mu^{00}$.} 
\label{tab:virgo}
\footnotesize
\begin{tabular}{lccc}
% ***********************************************
\hline
\hline
\noalign{\smallskip}
% ***********************************************
\multicolumn{1}{c}{}                         &
\multicolumn{1}{c}{UMa}                      & 
\multicolumn{1}{c}{Virgo}                    & 
\multicolumn{1}{c}{Fornax}                   \\
% ***********************************************
\noalign{\smallskip}
\hline
\noalign{\smallskip}
% ***********************************************
TRGB           & \nodata        & $>31.3^{1)}$   & \nodata        \\
Cepheids       & \nodata        & $31.45^{2)}\pm0.27$ & $31.53^{2)}\pm0.23$ \\
SNe\,Ia        & \nodata        & $31.54^{2)}\pm0.29$ & $31.60^{2)}\pm0.15$ \\
TF             & $31.45\pm0.06$ & $31.62\pm0.16$ & \nodata        \\
$D_{n}\!-\!\sigma$ & \nodata    & $31.60\pm0.10$ & $31.69\pm0.16$ \\
\noalign{\smallskip}
\hline
\noalign{\smallskip}
adopted (weighted) & $31.45\pm0.06$ & $31.59\pm0.08$ & $31.62\pm0.10$ \\
distance (Mpc)     &  $19.5\pm0.6$  &  $20.3\pm0.3$  &  $21.1\pm1.1$  \\
$v_{220}$          & $1253\;(\pm40)$& $1152\;(\pm35)$& $1371\;(\pm30)$\\  
$H_{0}$            &  $64.3\pm2.9$  &  $55.4\pm2.7$  &  $65.0\pm3.5$  \\
% ***********************************************
\noalign{\smallskip}
\hline
\noalign{\smallskip}
\multicolumn{4}{l}{\small $^{1)}$ \citeauthor*{TSR:08} from data of
  \citet{Durrell:etal:07} and \citet{Caldwell:06}} \\
\multicolumn{4}{l}{\small $^{2)}$ individually listed in 
\citeauthor*{STS:06}} \\
% ***********************************************
\end{tabular}
\end{center}
\end{table}
% ***********************************************

% ********************************************************
% Figure 4: Hubble diagram of 13 groups and clusters
% ********************************************************
% ********************************************************

% ******************************************************************
% 6. Conclusions
% ******************************************************************
\section{Conclusions}
\label{sec:6}
An intercomparison of RR\,Lyr, TRGB, and Cepheid distances shows that
their dispersion is not more than $0.15\mag$. The same upper limit
holds for SNe\,Ia as seen from the small scatter in their Hubble
diagram at large distances. The four distance indicators stand out
because they can provide the most accurate distances within their
reach for sizable samples of galaxies and, importantly, their small
dispersion makes them highly insensitive to selection bias. Although
their reach is drastically different, RR\,Lyr stars being very
short-range, SNe\,Ia extending to cosmological distances, and the TRGB
and Cepheid distances lying in between, there is enough overlap to tie
them into a single system of distances. 
 
     The combined Hubble diagram of TRGB, Cepheid, and SNe\,Ia
distances shows a well defined Hubble line with slope 0.2,
corresponding to linear expansion, over a range of $\sim\!250$ to at
least $20,000\kms$. The slope of 0.2, strongly supported also by other
evidence (see Sect.~\ref{sec:2}) implies that the present mean value
of the Hubble constant $H_{0}$ {\em is everywhere the same} 
(cosmological effects being exempt by definition). 
Most of the observed dispersion about the Hubble line must be caused
by random peculiar motions; allowing for the (small) distance errors
they are $70\kms$ within $7\;$Mpc and increase outwards to a
yet undetermined limit (see Table~\ref{tab:char}, Col.~10). Lower values
are in the literature, but the value here seems well determined from
78 TRGB distances  (Table~\ref{tab:char}). 

     The zero point of the Hubble line is set in two {\em independent\/}
ways. a) The absolute magnitude of the TRGB is determined by 22 RR\,Lyr
star distances and agrees well with other determinations. The adopted
magnitude of $M^{I}_{\rm TRGB}=-4.05$ carries a systematic error of
hardly more than $0.1\mag$. 
The value holds for [Fe/H]$_{\rm ZW}=-1.6$ and changes by less than
$0.1\mag$ in the range $-2.0<\mbox{[Fe/H]}_{\rm ZW}<-1.3$ typical for
old populations (\citeauthor*{TSR:08}, Fig.~1).
The resulting value of $H_{0}=63.0\pm1.6\;(\pm3)$ from 78 distances
larger than $4.5\;$Mpc refers to an effective distance of only
$\sim\!400\kms$ where the influence of peculiar velocities is still
large, but this is compensated by the large number of TRGB distances. 
b) Because the P-L relations of the metal-rich Galactic Cepheids and
of the metal-poor LMC Cepheids are different they are independently
calibrated. The zero point of the Galactic P-L relation rests on
Cepheids in Galactic clusters and on physical BBW distances. The 
zero point of a 10-day Cepheid is confirmed by trigonometric
parallaxes to within a few $0.01\mag$, but the error can increase 
to $\sim0.15\mag$ for Cepheids with the shortest and longest periods
depending on the correctness of the adopted P-L slope. 
The LMC P-L relation with very well determined slope is zero-pointed
by an adopted distance of $\mu^{0}_{\rm LMC} = 18.54$. This value is
based on a multitude of determinations, excluding of course results
depending on the P-L relations of Cepheids themselves; the error is
again estimated to be $0.1\mag$. It should be noted that significantly
smaller LMC distances come mostly from some a priory assumption on the
shape and zero point of the P-L relations of the Galaxy and LMC. -- 
The Cepheids in other galaxies with metallicities like the Galaxy or
LMC are reduced with the corresponding P-L relations; in case of
Cepheids with intermediate metallicities the results from the two P-L
relations are interpolated. The resulting 29 Cepheid distances larger
than $4.5\;$Mpc give $H_{0}=63.4\pm1.7$ at an effective distance of
$900\kms$. --  
The good agreement of the value of $H_{0}$ from the TRGB and Cepheid
distances is highly significant because it is {\em predicted\/} from
the well supported linearity of the expansion field.
 
     SNe\,Ia are calibrated through Cepheids and cannot independently
contribute to the zero point of the distance scale. But their large
number can reduce the statistical error and serve to carry the value 
of $H_{0}$ to $\sim\!20\,000\kms$. They give $H_{0}=60.3\pm2.6$ at an
effective distance of $1600\kms$ and, allowing for a flat Universe
with $\Omega_{\Lambda}=0.7$, $H_{0}=61.2\pm0.5$ from 62 SNe\,Ia at 
$v>3000\kms$.  
The adopted value of 
\begin{equation}
H_{0} = 62.3\pm1.3\quad (\pm4.0) 
\label{eq:H0adopt}
\end{equation}
is the unweighted mean from the Cepheids and Cepheid-calibrated SNe\,Ia
averaged with the result from the TRGB. The generous 6\% systematic
error is estimated in Sect.~\ref{sec:3:4:2}.  

     The value of $H_{0}$ rests on the two independent zero points set
by the TRGB and Cepheid distances. No other zero-pointed distance
indicator is available at present, which could carry the distance
scale into the expansion field, i.e.\ to $>4.5\;$Mpc, for a sufficient
number of 20 or more galaxies. But TF distances of a distance-limited
sample of spiral galaxies and $D_{n}\!-\!\sigma$ distances out to 
$2500\kms$ as well the Hubble diagram of nearby groups and clusters
provide at least a consistency check. We are not aware of any serious
objection against the adopted value of $H_{0}$.
 
     The literature abounds in larger values of $H_{0}$. Some are
based on the untenable view that the LMC P-L relation of Cepheids,
whatever its exact shape and zero point, is universal. Others are the
result of selection bias, which becomes particularly severe when it is
tried to determine $H_{0}$ at the largest distances which can be
reached, and from where necessarily only the most luminous objects of
their species can enter the catalogs. 
The importance of selection bias is often underestimated because the
quality of the distance indicators is overestimated. The true quality
can be determined only if there is broad overlap with high-accuracy
distance indicators like RR\,Lyr stars or TRGB and Cepheid distances,
or by consulting the Hubble diagram. The dispersion here, corrected
for the reasonably well understood effect of peculiar velocities,
gives the random error for a given distance indicator. Also too steep
a slope, i.e.\ $H_{0}$ increasing with distance, is a clear sign of
important bias or some other systematic problem of the method. Finally
other high values of $H_{0}$ are too model-dependent to be reliable.  

     Future progress on $H_{0}$ will come from additional
near-infrared photometry of Cepheids where they are relatively
insensitive to absorption and metallicity. Enormous potential lies
still in the TRGB distances. With a somewhat improved understanding of
their metallicity dependence, which is in any case small in old
populations, they can provide distances to better than $\pm5\%$ for
well over 1000 galaxies of all types within $\sim\!20\;$Mpc with
present techniques and requiring relatively little telescope
time. They will thus map the local velocity field in great detail and
also yield a high-weight calibration of SNe\,Ia extending the impact
of the method to cosmological distances.

% ***********************************************
%  Acknowledgments
% ***********************************************
\begin{acknowledgements}
We thank our many collaborators over the years, particularly 
Abhijit Saha, for their work with us on $H_{0}$. 
A.\,S. thanks the Observatories of the Carnegie Institution for
post-retirement facilities.
\end{acknowledgements}

% ******************************************************************
% Bibliography
% ******************************************************************

% ******************************************************************

% ******************************************************************
\end{document}